\documentclass[12pt]{article}

\usepackage{amsfonts}
\usepackage{latexsym}
\usepackage{amsmath}
\usepackage{amssymb}
\usepackage{color}

\begin{document}

 \title{Inf-convolution of G-expectations}

\date{}
\author{Xuepeng Bai\thanks{Corresponding author }\ \ \ Rainer Buckdahn$^\dag$}
 \maketitle
\noindent \vspace{-1.6cm} \begin{flushleft} \hspace{.75cm}
 $^*$\scriptsize{School of Mathematics, Shandong
University, 250100 Jinan, P.R China}\\\hspace{1cm}\scriptsize{Email:
xuepeng.bai@gmail.com}
\\\hspace{.85cm}$^\dag$\scriptsize{Universit$\acute{e}$ de Bretagne Occidentale, Laboratoire de Math\'ematiques, CNRS-UMR 6205, \\ \hspace{1cm}F-29238 BREST Cedex 3, France}
\\\hspace{1cm}\scriptsize{Email: Rainer.Buckdahn@univ-brest.fr}
\end{flushleft}

 \abstract{In this paper we will discuss the optimal risk transfer problems when risk measures are generated by G-expectations,
 and we present the relationship between inf-convolution of G-expectations and the inf-convolution of drivers G.}
\section{Introduction}

Coherent risk measures were introduced by Artzner et al. [1] in
finite probability spaces and lately by Delbaen [8,9] in general
probability spaces. The family of coherent risk measures was
extended later by F\"{o}llmer and Schied [10,11] and, independently,
by Frittelli and Rosazza Gianin [12,13] to the class of convex risk
measures.

The notion of g-expectations was introduced by Peng [15] as
solutions to a class of nonlinear Backward Stochastic Differential
Equations (BSDE in short) which were first studied by Pardoux and
Peng [14]. Financial applications were discussed in detail by El
Karoui et al. [6].

Let us introduce the optimal risk transfer model we are concerned
with. This model can be briefly described as follows:

Two economic agents A and B are considered, who assess the risk
associated with their respective positions by risk measures $\rho_A$
and $\rho_B$. The issuer, agent A, with the total risk capital X,
wants to issue a financial product F and sell it to agent B for the
price $\pi$ in order to reduce his risk exposure. His objective is
to minimize $\rho_A(X-F+\pi)$ with respect to F and $\pi$, while the
interest of buyer B is not to be exposed to a greater risk after the
transaction:
$$\rho_B(F-\pi)\leq \rho_B(0).$$
Using the cash translation invariance property, this optimization
problem can be rewritten in the simpler form
$$\inf_F\{\rho_A(X-F)+\rho_B(F)\}.$$

This problem was first studied by El Karoui and Pauline Barrieu
[2,3,4] for convex risk measures, in particular those described by
g-expectation.

Related with the pioneering paper [1] on coherent risk measures,
sublinear expectations (or, more generally, convex expectations, see
[10,11,13]) have become more and more popular for modeling such risk
measures. Indeed, in any sublinear expectation space $(\Omega,{\cal
H},\hat{\mathbb{E}})$ a coherent risk measure $\rho$ can be defined
in a simple way by putting $\rho(X):=\hat{\mathbb{E}}[-X],$ for
$X\in{\cal H}$.

The notion of a sublinear expectation named G-expectation was first
introduced by Peng [17,18] in 2006. Compared with g-expectations,
the theory of G-expectation is intrinsic in the sense that it is not
based on a given (linear) probability space. A G-expectation is a
fully nonlinear expectation. It characterizes the variance
uncertainty of a random variable. We recall that the problem of mean
uncertainty has been studied by Chen-Epstein through g-expectation
in [5]. Under this fully nonlinear G-expectation, a new type of
It$\hat{\mbox{o}}$'s formula has been obtained, and the existence
and uniqueness for stochastic differential equation driven by a
G-Brownian motion have been shown. For a more detailed description
the reader is  referred to Peng's recent papers [17,18,19].

This paper focuses on the mentioned optimization problem where the
g-risk measures are replaced by one dimensional G-expectations,
i.e., the problem:
$$\hat{\mathbb{E}}_{G_1}\square\ \hat{\mathbb{E}}_{G_2}[X]:= \inf_F\{\hat{\mathbb{E}}_{G_1}[X-F]+\hat{\mathbb{E}}_{G_2}[F]\}.$$
The main aim of this paper is to present the relationship between
the above introduced operator $\hat{\mathbb{E}}_{G_1}\square\
\hat{\mathbb{E}}_{G_2}[\cdot]$ and the G-expectation
$\hat{\mathbb{E}}_{{G_1}\square {G_2}}[\cdot]$. More precisely, we
show that both operators coincide if ${G_1}\square
{G_2}\neq-\infty.$

In this paper we constrain ourselves to one dimensional
G-expectation, the multi-dimensional case is much more complicated
and we hope to study this case in a forthcoming publication.

Our approach is mainly based on the recent results by Peng [19]
which allow to show that $\hat{\mathbb{E}}_{G_1}\square\
 \hat{\mathbb{E}}_{G_2}[\cdot]$ constructed by inf-convolution of
 $\hat{\mathbb{E}}_{G_1}[\cdot]$ and $\hat{\mathbb{E}}_{G_2}[\cdot]$
 satisfies  the properties of  G-expectation. To our best knowledge, this is the first
 paper that uses the results of Theorem 4.1.3 of [19] to prove that a given nonlinear
 expectation is a G-expectation.

This paper is organized as follows: while basic definitions and
properties of G-expectation and G-Brownian Motion are recalled in
Section 2, Section 3 states and proves the main result of this
paper: If ${G_1}\square {G_2}\neq-\infty$, then
$\hat{\mathbb{E}}_{G_1}\square\ \hat{\mathbb{E}}_{G_2}[\cdot]$ also
is a G-expectation and $$\hat{\mathbb{E}}_{G_1}\square\
\hat{\mathbb{E}}_{G_2}[\cdot]=\hat{\mathbb{E}}_{{G_1}\square
{G_2}}[\cdot].$$

 \section{ Notation and
Preliminaries}

The aim of this section is to recall some basic definitions and
properties of G-expectations and G-Brownian motions, which will be
needed in the sequel. The reader interested in a more detailed
description of these notions is referred to Peng's recent papers
[17,18,19]. \\
Adapting Peng's approach in [19], we let
 $\Omega$ be a given  nonempty fundamental space and $\mathcal {H}$ be a linear space of real
functions defined on $\Omega$ such that :\\\\
$ \quad i)$ $1\in \mathcal {H}$. \\
$\quad ii)$ $\mathcal {H}$ is stable with respect to local Lipschitz
functions, i.e. for all $n\geq1$, and for all $X_1,...,X_n\in
\mathcal {H}$, $\varphi \in C_{l,lip}(\mathbb{R}^n),$ it holds also
$\varphi(X_1,...,X_n)\in\mathcal {H}$.\\\\
Recall that  $ C_{l,lip}(\mathbb{R}^n)$ denotes the space of all
local Lipschitz functions $\varphi$ over $\mathbb{R}^n$ satisfying
$$|\varphi(x)-\varphi(y)|\leq C(1+|x|^m+|y|^m)|x-y|, x,y \in
\mathbb{R}^n,$$ for some $C>0,m\in\mathbb{N}$ depending on
$\varphi$. The set $\mathcal {H}$ is interpreted as the space of
random variables
defined on $\Omega$.\\
\\\textbf{Definition 2.1}\ A sublinear expectation
$\hat{\mathbb{E}}$ on $\mathcal {H}$ is  a functional $\mathcal
{H}\rightarrow\mathbb{R}$ with the following properties : for
all $X,Y\in \mathcal {H}$, we have\\
\textbf{(a) Monotonicity:} if $X\geq Y$
then $\hat{\mathbb{E}}[X]\geq\hat{\mathbb{E}}[Y].$\\
\textbf{(b) Preservation of constants:}\ $\hat{\mathbb{E}}[c]=c,$ for all reals c. \\
\textbf{(c) Sub-additivity (or property of self-dominacy)}:
$$\hat{\mathbb{E}}[X]-\hat{\mathbb{E}}[Y]\leq\hat{\mathbb{E}}[X-Y].$$
\textbf{(d) Positive homogeneity:}\ $\hat{\mathbb{E}}[\lambda
X]=\lambda\hat{\mathbb{E}}[X],\forall \lambda\geq 0.$
\\
\\
The triple $(\Omega,\mathcal {H},\hat{\mathbb{E}})$  is called a
sublinear expectation space. It generalizes the classical case of
the linear expectation $E[X]=\int_\Omega X dP,\ X \in
L^1(\Omega,\mathcal {F},\mathcal {P}),$ over a probability space
$(\Omega,\mathcal{F},\mathcal {P}).$ Moreover,
 $\rho(X)=\hat{\mathbb{E}}[-X]$ defines
a coherent risk measure on $\mathcal {H}$.
\\\\
\textbf{Definition 2.2} For arbitrary $n,m \geq 1,$ a random vector
$Y=(Y_1,Y_2,...,Y_n)\in\mathcal {H}^n\ (=\mathcal
{H}\times\mathcal{H}\times ...\times \mathcal {H})$ is said to be
independent of $X\in\mathcal {H}^m$ under $\hat{\mathbb{E}}[\cdot]$
if for each test function $\varphi\in C_{l,lip}(\mathbb{R}^{n+m}) $
we have
$$\hat{\mathbb{E}}[\varphi(X,Y)]=\hat{\mathbb{E}}[\hat{\mathbb{E}}[\varphi(x,Y)]_{x=X}].$$\\
\textbf{Remark:} In the case of linear expectation, this notion of
independence is just the classical one. It is important to note that
under sublinear expectations the condition Y is independent to X
does not imply automatically that X is independent to Y.\\\\
 Let $X=(X_1,...,X_n)\in\mathcal {H}^{ n}$ be a
given random vector. We define a functional on
$C_{l,lip}(\mathbb{R}^n)$ by
$$\hat{\mathbb{F}}_X[\varphi]:=\hat{\mathbb{E}}[\varphi(X)],\varphi\in
C_{l,lip}(\mathbb{R}^n).$$ It's easy to check that
$\hat{\mathbb{F}}_X[\cdot]$ is a sublinear expectation defined
on $(\mathbb{R}^n,C_{l,lip}(\mathbb{R}^n))$.\\\\
 \textbf{Definition
2.3}  Given two sublinear expectation spaces $(\Omega,\mathcal
{H},\hat{\mathbb{E}})$ and $(\widetilde{\Omega},\widetilde{\mathcal
{H}},\widetilde{\mathbb{E}})$, two random vectors $X\in\mathcal
{H}^{n}$ and $Y\in\widetilde{\mathcal {H}}^{n}$  are said to be
identically distributed if for each test function $\varphi\in
C_{l,lip}(\mathbb{R}^n)$
$$\hat{\mathbb{F}}_X[\varphi]=\widetilde{\mathbb{F}}_Y[\varphi].$$\\
We now introduce the important notion of G-normal distribution. For
this, let $0\leq\underline{\sigma}\leq\overline{\sigma} \in
\mathbb{R}$, and let G be the sublinear function:
$$G(\alpha)=\frac{1}{2}(\overline{\sigma}^2\alpha^+-\underline{\sigma}^2\alpha^-),\alpha\in
\mathbb{R}.$$ As usual $\alpha^+=max\{0,\alpha\}$ and
$\alpha^-=(-\alpha)^+.$ Given an arbitrary initial condition
$\varphi \in C_{l,lip}(\mathbb{R})$, we denote by $u_\varphi$ the
unique viscosity solution of the following parabolic partial
differential equation (PDE):
\begin{eqnarray*}
&&\partial_t u_\varphi(t,x)=G(\partial_{xx}^2 u_\varphi(t,x)),\ \ \
\ \ \ \
(t,x)\in (0,\infty)\times\mathbb{R},\\
&&u_\varphi(0,x)=\varphi(x),\ \ \ \ \ \ \ \ \ \ \ \ \ \ \ \ \ \ \ \
\ \ \ \ \ x\in\mathbb{R}.
\end{eqnarray*}\\
 \textbf{Definition 2.4 :} A random
variable X in a sub-expectation space $(\Omega,\mathcal
{H},\hat{\mathbb{E}})$ is called
$G_{\underline{\sigma},\overline{\sigma}}$-normal distributed, and
we write $X\sim\mathcal
{N}(0;[\underline{\sigma}^2,\overline{\sigma}^2])$, if for all
$\varphi\in C_{l,lip}(\mathbb{R}),$
$$\hat{\mathbb{E}}[\varphi(x+\sqrt{t}X)]:=u_\varphi(t,x),\ \ \ \ \ \ \ \ (t,x)\in
[0,\infty)\times\mathbb{R}.$$
\\
\textbf{Remark:} From [18], we have the following Kolmogrov-Chapman
chain rule:
$$u_\varphi(t+s,x)=\hat{\mathbb{E}}[u_{\varphi}(t,x+\sqrt{s}X)],\ \ \ s\geq 0.$$
\\In what follows we will take as fundamental space $\Omega $ the
space $C_0(\mathbb{R}^+)$ of all real-valued continuous functions
$(\omega_t)_{t\in\mathbb{R}^+}$ with $\omega_0=0$, equipped with the
topology generated by the uniform convergence on compacts.\\ For
each fixed $ T\geq0$, we consider the following space of local
Lipschitz functionals :\\
\[\begin{tabular}{l}$\mathcal {H}_T=Lip(\mathcal {F}_T):$\\
$\ \ \ \ \ =\{X(\omega)=\varphi(\omega_{t_1},...,\omega_{t_m}),
t_1,...,t_m\in [0,T],\varphi \in
C_{l,lip}(\mathbb{R}^m),m\geq1\}$.\end{tabular}\] Furthermore, for
$0\leq s \leq t,$ we define
\[\begin{tabular}{l}$\mathcal{H}_t^s=Lip(\mathcal{F}_t^s):$\\
$\ \ \ \ \
=\{X(\omega)=\varphi(\omega_{t_2}-\omega_{t_1},...,\omega_{t_{m+1}}-\omega_{t_{m}}),
t_1,...,t_{m+1}\in [s,t],$\\$\ \ \ \ \ \ \ \ \ \ \  \varphi \in
C_{l,lip}(\mathbb{R}^m),m\geq1\}.$\end{tabular}\] It is clear that
$\mathcal{H}_t^s\subseteq\mathcal {H}_t\subseteq Lip(\mathcal
{F}_T),\ \mbox{for}\ s\leq t\leq T.$ We also introduce the space
$$\mathcal {H}=Lip(\mathcal {F}):=\bigcup^\infty_{n=1}Lip(\mathcal
{F}_n).$$\\
Obviously, $Lip(\mathcal{F}_t^s),\ Lip(\mathcal {F}_T) \ \mbox{and}\
Lip(\mathcal {F})$ are vector lattices.
\\\\
We will consider the canonical space and set
$$B_t(\omega)=\omega_t,t\in[0,\infty),\ \mbox{for}\ \omega\in\Omega.$$
Obviously, for each $t\in [0,\infty),B_t\in Lip(\mathcal {F}_t).$
Let
$G(a)=G_{\underline{\sigma},\overline{\sigma}}(a)=\frac{1}{2}(\overline{\sigma}^2a^+-\underline{\sigma}^2a^-),a\in
\mathbb{R}.$ We now introduce a sublinear expectation
$\hat{\mathbb{E}}$ defined on $\mathcal {H}_T=Lip(\mathcal {F}_T),$
as well as on $\mathcal {H}=Lip(\mathcal {F}),$ via the following
procedure: For each $X\in \mathcal {H}_T$ with
$$X=\varphi(B_{t_1}-B_{t_0},B_{t_2}-B_{t_1},...,B_{t_m}-B_{t_{m-1}}),$$
and for all $\varphi\in C_{l,lip}(\mathbb{R}^m)$ and $0=t_0\leq
t_1<...<t_m\leq T,\ m\geq 1,$ we set
\begin{eqnarray*}
&&\hat{\mathbb{E}}[\varphi(B_{t_1}-B_{t_0},B_{t_2}-B_{t_1},...,B_{t_m}-B_{t_{m-1}})]\\
&&=\widetilde{\mathbb{E}}[\varphi(\sqrt{t_1-t_0}\xi_1,...,\sqrt{t_m-t_{m-1}}\xi_m)],
\end{eqnarray*}
where $(\xi_1,...,\xi_m)$ is an m-dimensional random vector in some
sublinear expectation space $(\widetilde{\Omega},\widetilde{\mathcal
{H}},\widetilde{\mathbb{E}}),$ such that $\xi_i\sim\mathcal
{N}(0;[\underline{\sigma}^2,\overline{\sigma}^2])$ and $\xi_{i+1}$
is independent of $(\xi_1,...,\xi_i),$ for all $i=1,...,m-1, m\in
\mathbb{N}.$ The related conditional expectation of
$X=\varphi(B_{t_1}-B_{t_0},B_{t_2}-B_{t_1},...,B_{t_m}-B_{t_{m-1}})$
under $\mathcal {H}_{t_j}$ is defined by
\begin{eqnarray*}
&&\hat{\mathbb{E}}[X|\mathcal
{H}_{t_j}]=\hat{\mathbb{E}}[\varphi(B_{t_1}-B_{t_0},B_{t_2}-B_{t_1},...,B_{t_m}-B_{t_{m-1}})|\mathcal
{H}_{t_j}]\\
&&\ \ \ \ \ \ \ \ \ \ \ \ \
=\psi(B_{t_1}-B_{t_0},...,B_{t_j}-B_{t_{j-1}})
\end{eqnarray*}
where
$$\psi(x_1,...,x_j)=\widetilde{\mathbb{E}}[\varphi(x_1,...,x_j,\sqrt{t_{j+1}-t_j}\xi_{j+1},...,\sqrt{t_m-t_{m-1}}\xi_m)].$$
We know from [18,19] that $\hat{\mathbb{E}}[\cdot]$ defines
consistently
 a sublinear expectation on $Lip(\mathcal {F}),$
satisfying (a)-(d) in Definition 2.1. The reader interested in a more detailed discussion is referred to [18,19]. \\ \\
\textbf{Definition 2.5} The expectation
$\hat{\mathbb{E}}[\cdot]:Lip(\mathcal {F})\rightarrow\mathbb{R}$
defined through the above procedure is called
$G_{\underline{\sigma},\overline{\sigma}}$-expectation. The
corresponding canonical process $(B_t)_{t\geq0}$ in the sublinear
expectation is called  a
$G_{\underline{\sigma},\overline{\sigma}}$-Brownian motion on $(\Omega,\mathcal {H},\hat{\mathbb{E}})$. \\
At the end of this section we list some useful properties that we
will need in Section 3.\\\\
 \textbf{Proposition 2.6 ([18,19])} The  following properties of
 $\hat{\mathbb{E}}[\cdot|\mathcal {H}_t]$  hold for all $X,Y\in\mathcal
 {H}=Lip(\mathcal {F}):$\\
 (a')If $X\geq Y$, then $\hat{\mathbb{E}}[X|\mathcal
 {H}_t]\geq\hat{\mathbb{E}}[Y|\mathcal {H}_t].$\\
 (b')$\hat{\mathbb{E}}[\eta|\mathcal {H}_t]=\eta,$ for each
 $t\in[0,\infty)$ and $\eta\in \mathcal {H}_t$.\\
 (c')$\hat{\mathbb{E}}[X|\mathcal {H}_t]-\hat{\mathbb{E}}[Y|\mathcal
 {H}_t]\leq \hat{\mathbb{E}}[X-Y|\mathcal {H}_t].$\\
 (d')$\hat{\mathbb{E}}[\eta X|\mathcal
 {H}_t]=\eta^+\hat{\mathbb{E}}[X|\mathcal {H}_t]+\eta^-\hat{\mathbb{E}}[-X|\mathcal
 {H}_t],$ for each $\eta\in \mathcal {H}_t$.\\
 We also have $$\hat{\mathbb{E}}[\hat{\mathbb{E}}[X|\mathcal {H}_t]|\mathcal
 {H}_s]=\hat{\mathbb{E}}[X|\mathcal {H}_{t\wedge s}],\ \mbox{and in\
 particular,}\
 \hat{\mathbb{E}}[\hat{\mathbb{E}}[X|\mathcal
 {H}_t]]=\hat{\mathbb{E}}[X].$$
 For each $X\in Lip(\mathcal {F}_T^t),\ \hat{\mathbb{E}}[X|\mathcal
 {H}_t]=\hat{\mathbb{E}}[X],$
moreover, the properties (b') and (c') imply:
$\hat{\mathbb{E}}[X+\eta|\mathcal {H}_t]=\hat{\mathbb{E}}[X|\mathcal
{H}_t]+\eta,$ whenever $\eta\in
\mathcal {H}_t.$\\
We will need  also the following two propositions, and for proofs
the
reader is referred to [18,19].\\
\\
\textbf{Proposition 2.7} For each convex function $\varphi$ and each
concave function $\psi$ with $\varphi(B_t)$ and $\psi(B_t)$ $\in
\mathcal {H}_t$, we have
$\hat{\mathbb{E}}[\varphi(B_t)]=\mathbb{E}[\varphi(\overline{\sigma}W_t)]$
and
$\hat{\mathbb{E}}[\psi(B_t)]=\mathbb{E}[\psi(\underline{{\sigma}}W_t)]$,
where
$(W_t)_{t\geq0}$ is a Brownian motion under the linear expectation $\mathbb{E}$.\\\\
\textbf{Proposition 2.8} Let $\hat{\mathbb{E}}_1[\cdot]$ and
$\hat{\mathbb{E}}_2[\cdot]$ be
 a $G_{\underline{\sigma}_1,\overline{\sigma}_1}$ and a
$G_{\underline{\sigma}_2,\overline{\sigma}_2}$ expectation on the
space $(\Omega,\mathcal {H}),$ respectively. Then, if
$[\underline{\sigma}_1,\overline{\sigma}_1]\subseteq[\underline{\sigma}_2,\overline{\sigma}_2],$
 we have $\hat{\mathbb{E}}_1[X]\leq\hat{\mathbb{E}}_2[X]$ and $\hat{\mathbb{E}}_1[X|\mathcal {H}_t]\leq\hat{\mathbb{E}}_2[X|\mathcal {H}_t]$,  for all $X\in
\mathcal {H}$ and all $t\geq 0.$

\section{ Inf-convolution of G-expectations}

The aim of this section is to state the main result of this paper,
that is the relationship between the inf-convolution
$\hat{\mathbb{E}}_{G_1}\square\ \hat{\mathbb{E}}_{G_2}[\cdot]$ and
the G-expectation $\hat{\mathbb{E}}_{{G_1}\square {G_2}}[\cdot]$. We
begin with the definitions necessary for the understanding of these both expressions. \\
For given $0\leq\underline{\sigma}_i\leq\overline{\sigma}_i\in
\mathbb{R}$, i=1,2, let
$G_i=G_{\underline{\sigma}_i,\overline{\sigma}_i}$ and we denote by
$\hat{\mathbb{E}}_i[\cdot ]$ the $G_i$-expectation
$\hat{\mathbb{E}}_{G_i}[\cdot ]$ on $(\Omega,\mathcal {H})$
$(=(C_0(\mathbb{R}^+),Lip(\mathcal {F}))).$ The inf-convolution of
$\hat{\mathbb{E}}_1[\cdot]$ with $\hat{\mathbb{E}}_2[\cdot]$,
denoted by $\hat{\mathbb{E}}_1\square\ \hat{\mathbb{E}}_2[\cdot] $
is defined as :
$$\hat{\mathbb{E}}_1\square\ \hat{\mathbb{E}}_2[X]= \inf_{F\in\mathcal
{H}}\{\hat{\mathbb{E}}_1[X-F]+\hat{\mathbb{E}}_2[F]\},\ \ \  X\in
\mathcal {H}.$$ Notice that $\hat{\mathbb{E}}_1\square\
\hat{\mathbb{E}}_2[\cdot ]:\ \ \mathcal {H}\rightarrow
\mathbb{R}\cup
\{-\infty\}.$ \\
In the same way we define
$$G_1\Box G_2(x)=\inf_{y\in \mathbb{R}}\{G_1(x-y)+G_2(y)\},\ \  x\in
\mathbb{R}.$$ Observe also that $G_1\Box G_2(\cdot):\
\mathbb{R}\rightarrow\ \mathbb{R}\cup\{-\infty\}.$ It is easy to
check that $G_1\Box G_2(\cdot)$ has the following form:
\[G_1\Box G_2(x)=\left\{\begin{tabular}{ll}$-\infty,$\ \ \ \ \
&$[\underline{\sigma}_1,\overline{\sigma}_1]\cap[\underline{\sigma}_2,\overline{\sigma}_2]=\emptyset;$\\
 $\frac{1}{2}(\overline{\sigma}^2x^+-\underline{\sigma}^2x^-),$ \ \ \
  &$[\underline{\sigma}_1,\overline{\sigma}_1]\cap[\underline{\sigma}_2,\overline{\sigma}_2]=[\underline{\sigma},\overline{\sigma}]\neq\emptyset.$
 \end{tabular} \right.\]
If $ G_1\Box G_2(\cdot)=-\infty$, then also
$\hat{\mathbb{E}}_1\square\ \hat{\mathbb{E}}_2[\cdot]=-\infty$. More
precisely, we have the following proposition:\\\\
 \textbf{Proposition 3.1} If
 $[\underline{\sigma}_1,\overline{\sigma}_1]\cap[\underline{\sigma}_2,\overline{\sigma}_2]=\emptyset$,
 then $\hat{\mathbb{E}}_1\square\hat{\mathbb{E}}_2[X]=-\infty$, for all $ X\in\mathcal
 {H}.$\\
 Proof:  Without loss of generality we may suppose
 $\overline{\sigma}_1<\underline{\sigma}_2.$ Choosing $F=-\lambda B_t^2,\lambda>0,t>0,$  we then
 have due to Proposition 2.7 that for all $X\in\mathcal {H},$
\begin{eqnarray*}
&&\ \ \ \hat{\mathbb{E}}_1[X-F]+\hat{\mathbb{E}}_2[F]\\
&&=\hat{\mathbb{E}}_1[X+\lambda
B_t^2]+\hat{\mathbb{E}}_2[-\lambda B_t^2]\\
&&\leq \hat{\mathbb{E}}_1[X]+\hat{\mathbb{E}}_1[\lambda
B_t^2]+\hat{\mathbb{E}}_2[-\lambda B_t^2]\\
&&\leq
\hat{\mathbb{E}}_1[X]+\lambda\overline{\sigma}_1^2t-\lambda\underline{\sigma}_2^2t.
\end{eqnarray*}
Letting $\lambda\rightarrow\infty,$ we obtain $ \hat{\mathbb{E}}_1\square\ \hat{\mathbb{E}}_2[X]=-\infty.$\ \ \ \ \ \ \
\ \ \ \ \ \ \ \ \ \ \ \ \ \ \ \ \ \ \ \ \ \ \ \ \ \ \ \ \ \ \ \ \ \ \ \ \ \ \ \ \
 \  \ \ \ \ $\blacksquare$ \\
\\
If
$[\underline{\sigma}_1,\overline{\sigma}_1]\cap[\underline{\sigma}_2,\overline{\sigma}_2]$
is not empty we have the following theorem, which is the main result
of this paper.
\\  \\
\textbf{Theorem 3.2}\ \ Let $\hat{\mathbb{E}}_1[\cdot]$ and
$\hat{\mathbb{E}}_2[\cdot]$ be the two G-expectations on the space
$(\Omega,\mathcal {H}),$ which have been defined above. If $G_1\Box
G_2(\cdot)\neq -\infty$, then $\hat{\mathbb{E}}_1\square\
\hat{\mathbb{E}}_2[\cdot]$ is a G-expectation on $(\Omega,\mathcal
{H})$ and has the driver $G_1\Box G_2,$ i.e.,
$\hat{\mathbb{E}}_1\square\
\hat{\mathbb{E}}_2[\cdot]=\hat{\mathbb{E}}_{G_1\square G_2}[\cdot].$\\\\
Let us first discuss Theorem 3.2 in the special case.\\\\
\textbf{Lemma 3.3} Let
$[\underline{\sigma}_1,\overline{\sigma}_1]\subseteq[\underline{\sigma}_2,\overline{\sigma}_2].$
Then
  $G_1\Box G_2(\cdot)=G_1(\cdot)$, as well as
$\hat{\mathbb{E}}_1\square\
\hat{\mathbb{E}}_2[\cdot]=\hat{\mathbb{E}}_1[\cdot].$\\ \\
Proof: We already know that $G_1\Box G_2(\cdot)=G_1(\cdot),$ so it
remains only to prove that $\hat{\mathbb{E}}_1\square\
\hat{\mathbb{E}}_2[\cdot]=\hat{\mathbb{E}}_1[\cdot].$ For this we
note that, firstly, by choosing $F=0$ in the definition of
$\hat{\mathbb{E}}_1\square\ \hat{\mathbb{E}}_2,$ we get
$\hat{\mathbb{E}}_1\square\
\hat{\mathbb{E}}_2\leq\hat{\mathbb{E}}_i,\ i=1,2.$\\
On the other hand, due to Proposition 2.8 we know that
$\hat{\mathbb{E}}_1\leq\hat{\mathbb{E}}_2.$ Thus, from the
subadditivity of $\hat{\mathbb{E}}_1[\cdot],$
$$\hat{\mathbb{E}}_1[X-F]+\hat{\mathbb{E}}_2[F]\geq\hat{\mathbb{E}}_1[X-F]+\hat{\mathbb{E}}_1[F]\geq\hat{\mathbb{E}}_1[X],\ F\in\mathcal {H}.$$
Consequently,
$\hat{\mathbb{E}}_1\square\hat{\mathbb{E}}_2[\cdot]=\hat{\mathbb{E}}_1[\cdot]$.
Thus, Theorem 3.2 holds true in this special case.\\
The case
$[\underline{\sigma}_1,\overline{\sigma}_1]\supseteq[\underline{\sigma}_2,\overline{\sigma}_2]$
can be treated analogously.\ \ \ \ \ \ \  \ \ \ \ \ \ \ \ \ \ \ \ \
\ \ \ \ \ \ \ \ \ \ \ \ \ \ \ \ \ \ \  $\blacksquare$
\\\\
The situation becomes more complicate if neither
 $[\underline{\sigma}_1,\overline{\sigma}_1]
\subseteq[\underline{\sigma}_2,\overline{\sigma}_2]$ nor
$[\underline{\sigma}_2,\overline{\sigma}_2] \subseteq
[\underline{\sigma}_1,\overline{\sigma}_1].$ Without loss of
generality, we suppose that
$[\underline{\sigma}_1,\overline{\sigma}_1]\bigcap[\underline{\sigma}_2,\overline{\sigma}_2]\\=[\underline{\sigma}_2,\overline{\sigma}_1].$
In this case $$G_1\Box
G_2(x)=\frac{1}{2}(\overline{\sigma}_1^2x^+-\underline{\sigma}_2^2x^-)=G_3(x),\
x\in \mathbb{R},$$ where $G_3=G_{\underline{\sigma}_2
,\overline{\sigma}_1}$. By $\hat{\mathbb{E}}_3[\cdot]$ we denote the
G-expectation on $(\Omega,\mathcal {H})$ with driver $G_3(\cdot)$.
The above notations will be kept for the rest of the paper. Our aim
is to prove that $\hat{\mathbb{E}}_1\square\
\hat{\mathbb{E}}_2[\cdot]=\hat{\mathbb{E}}_3[\cdot].$  \\ \\
The proof is based on Theorem 4.1.3 in Peng's paper [19]; this
theorem characterizes the intrinsic properties of G-Brownian motions
and
G-expectations. \\\\
\textbf{Lemma 3.4 ( see Theorem 4.1.3, Peng [19])} Let
$(\widetilde{B}_t)_{t\geq0}$ be a process defined in the
sub-expectation space $(\widetilde{\Omega},\widetilde{\mathcal
{H}},\widetilde{\mathbb{E}}) $ such that\\
(i) $\widetilde{B}_0=0;$\\
(ii) For each $t,s\geq0,$ the increment
$\widetilde{B}_{t+s}-\widetilde{B}_t$ has the same distribution as
$\widetilde{B}_s$ and is independent of
$(\widetilde{B}_{t_1},\widetilde{B}_{t_2},...,\widetilde{B}_{t_n})$,
for all
 $0\leq t_1,...,t_n\leq t, n\geq1.$ \\
(iii)
$\widetilde{\mathbb{E}}[\widetilde{B}_t]=\widetilde{\mathbb{E}}[-\widetilde{B}_t]=0,$
and
$\lim_{t\downarrow0}\widetilde{\mathbb{E}}[|\widetilde{B}_t|^3]t^{-1}=0.$\\
Then $(\widetilde{B}_t)_{t\geq0}$ is a
$G_{\underline{\sigma},\overline{\sigma} }$-Brownian motion with
$\overline{\sigma}^2=\widetilde{\mathbb{E}}[\widetilde{B}_1^2]$ and
$\underline{\sigma}^2=-\widetilde{\mathbb{E}}[-\widetilde{B}_1^2].$
\\\\
In the sequel, in order to prove Theorem 3.2 we will show that the
inf-convolution $\hat{\mathbb{E}}_1\square\
\hat{\mathbb{E}}_2[\cdot]$ is a sublinear expectation on
$(\Omega,\mathcal {H})$. This will make Lemma 3.4 applicable. More
precisely, we will show that under the sublinear expectation
$\hat{\mathbb{E}}_1\square\ \hat{\mathbb{E}}_2[\cdot]$ the canonical
process $(B_t)_{t\geq0}$ satisfies the assumptions of Lemma 3.4 for
$\overline{\sigma}=\overline{\sigma}_1,\underline{\sigma}=\underline{\sigma}_2.$
This has as consequence that $(B_t)_{t\geq0}$ is a
$G_{\underline{\sigma}_2,\overline{\sigma}_1}$-Brownian motion under
$\hat{\mathbb{E}}_1\square\ \hat{\mathbb{E}}_2[\cdot],$ and implies
that $\hat{\mathbb{E}}_1\square\
\hat{\mathbb{E}}_2[\cdot]=\hat{\mathbb{E}}_3[\cdot].$\\\\
 \textbf{Proposition 3.5} Under the assumption $[\underline{\sigma}_1,\overline{\sigma}_1]\bigcap[\underline{\sigma}_2,\overline{\sigma}_2]
 =[\underline{\sigma}_2,\overline{\sigma}_1],$ the inf-convolution
$\hat{\mathbb{E}}_1\square\ \hat{\mathbb{E}}_2[\cdot]$ is a
sublinear expectation on $(\Omega,\mathcal {H})$.\\
Proof: (a) Monotonicity: The monotonicity is an immediate
consequence of that of the G-expectation
$\hat{\mathbb{E}}_1[\cdot].$\\\\
(b) Preservation of constants: From the preservation of constants
property and the subadditivity of $\hat{\mathbb{E}}_1,$ we have
\begin{eqnarray*}
&&\ \ \ \hat{\mathbb{E}}_1\square\
\hat{\mathbb{E}}_2[c]\\
&&=\inf_{F\in\mathcal
{H}}\{\hat{\mathbb{E}}_1[c-F]+\hat{\mathbb{E}}_2[F]\}
\\
&&=c+\inf_{F\in\mathcal
{H}}\{\hat{\mathbb{E}}_1[-F]+\hat{\mathbb{E}}_2[F]\}
\\
&&\geq c+\inf_{F\in\mathcal
{H}}\{\hat{\mathbb{E}}_3[-F]+\hat{\mathbb{E}}_3[F]\}\\
 &&\geq c.
\end{eqnarray*}
The latter lines follow from the fact that
$\hat{\mathbb{E}}_3\leq\hat{\mathbb{E}}_i, i=1,2,$ and the
subadditivity of $\hat{\mathbb{E}}_3.$ Moreover, by taking F=0 in
the definition of $\hat{\mathbb{E}}_1\square\
\hat{\mathbb{E}}_2[c]$ we get the converse inequality. \\\\
(c) Sub-additivity: Given arbitrary fixed $X,Y \in \mathcal {H}$, in
virtue of the subadditivity of $\hat{\mathbb{E}}_1[\cdot]$ and
$\hat{\mathbb{E}}_2[\cdot]$, we have for all $F_1,F_2 \in \mathcal
{H}$
\begin{eqnarray*}
&&\hat{\mathbb{E}}_1[X-Y-F_1]+\hat{\mathbb{E}}_2[F_1]+\hat{\mathbb{E}}_1[Y-F_2]+\hat{\mathbb{E}}_2[F_2]\\
&&\geq \hat{\mathbb{E}}_1[X-(F_1+F_2)]+\hat{\mathbb{E}}_2[F_1+F_2].
\end{eqnarray*}
Consequently,
\begin{eqnarray*}
&&\ \ \ \ \ \ \hat{\mathbb{E}}_1\square\
\hat{\mathbb{E}}_2[X-Y]+\hat{\mathbb{E}}_1\square\
\hat{\mathbb{E}}_2[Y]\\
 &&=\inf_{F_1,F_2\in \mathcal
{H}}\{\hat{\mathbb{E}}_1[X-Y-F_1]+\hat{\mathbb{E}}_2[F_1]+\hat{\mathbb{E}}_1[Y-F_2]+\hat{\mathbb{E}}_2[F_2]\}\\
&&\geq\inf_{F_1,F_2\in \mathcal
{H}}\{\hat{\mathbb{E}}_1[X-F_1-F_2]+\hat{\mathbb{E}}_2[F_1+F_2]\}\\
&&=\hat{\mathbb{E}}_1\square\ \hat{\mathbb{E}}_2[X].
\end{eqnarray*}
(d)Finally, the positive homogeneity is an easy consequence of that
of $\hat{\mathbb{E}}_1[\cdot]$ and
 $\hat{\mathbb{E}}_2[\cdot]$.\ \ \ \ \ \ \ \ \ \ \ \ \ \ \ \
 \ \ \ \ \ \ \ \ \ \ \ \ \ \ \ \ \ \ \ \ \ \ \ \
 \ \ \ \ \ \ \ \ \ \ \ \ \ \ \ \ \ \ \ \ \ \ \ \ \ \
 \ \ \ \ \ \ \ \ \ \ \ \ \ \ \ \ \ \ \ \ \ \ \ \ \ \ \ \ \ \ \ \ \ $\blacksquare$  \\\\
 The following series of statements has as objective to prove that
 the canonical process $(B_t)_{t\geq0}$ satisfies under the sublinear
 expectation $\hat{\mathbb{E}}_1\square\
\hat{\mathbb{E}}_2[\cdot]$ the assumptions of Lemma 3.4.\\\\
 \textbf{Lemma 3.6:} Let $\varphi$ be a convex or concave function such that $\varphi(B_t)\in \mathcal {H},$ then $\hat{\mathbb{E}}_1\square\
\hat{\mathbb{E}}_2[\varphi(B_t)]=\hat{\mathbb{E}}_3[\varphi(B_t)].$
\\
Proof: We only prove the convex case, the proof for concave
$\varphi$ is analogous. If $\varphi$ is convex we have according to
Proposition 2.7 ,
$$\hat{\mathbb{E}}_3[\varphi(B_t)]=\mathbb{E}[\varphi(\overline{\sigma}_1W_t)]=\hat{\mathbb{E}}_1[\varphi(B_t)].$$
By Proposition 2.8 we know that
$\hat{\mathbb{E}}_i[\cdot]\geq\hat{\mathbb{E}}_3[\cdot],i=1,2,$ and
consequently, also $\hat{\mathbb{E}}_1\square\
\hat{\mathbb{E}}_2[\cdot]\geq\hat{\mathbb{E}}_3[\cdot].$\\
On the other hand, since obviously, $\hat{\mathbb{E}}_1\square\
\hat{\mathbb{E}}_2[\cdot]\leq \hat{\mathbb{E}}_1[\cdot],$  we get,
for convex functions $\varphi$, $\hat{\mathbb{E}}_1\square\
\hat{\mathbb{E}}_2[\varphi(B_t)]=\hat{\mathbb{E}}_3[\varphi(B_t)].$
Similarly we can prove the concave case.\ \ \ \ \ \ \ \ \ \ \ \ \ \
\ \ \ \ \ \ \ \ \ \ \ \ \ \ \ \ \ \ \ \ \ \ \ \ \ \ \ \ \ \ \ \ \ \
\ \ \ \ \ \ \ \ \ \ \ \ \ \ \ \ \ \ \ \ \ \ \ \ \ \ \ \ \ \ \ \ \ \
\ \ \ \ \ \ \ \ \ \ \ \ \ \ \ \ \ \ \ \ \ \ \ \ \ \ \ \ \ \ \ \ \ \
\ \ \ \ \ \ \ \ \ \ $\blacksquare$
\\\\
\textbf{Remark:} From Proposition 3.5 we know already that
$\hat{\mathbb{E}}_1\square\ \hat{\mathbb{E}}_2[\cdot]$ is a
sublinear expectation. This implies $\hat{\mathbb{E}}_1\square\
\hat{\mathbb{E}}_2[0]=0$. From Lemma 3.6, we have that
 $F^*=0$ is an optimal control when $\varphi$ is convex, while
the optimal control is  $F^*=\varphi(B_t)$ when $\varphi$ is
concave. Moreover,
$$\hat{\mathbb{E}}_1\square \hat{\mathbb{E}}_2[-B_t]=\hat{\mathbb{E}}_1\square
\hat{\mathbb{E}}_2[B_t]=0$$
$$\hat{\mathbb{E}}_1\square\
\hat{\mathbb{E}}_2[B_t^2]=\overline{\sigma}_1^2t, \
\hat{\mathbb{E}}_1\square\
\hat{\mathbb{E}}_2[-B_t^2]=-\underline{\sigma}_2^2t.$$\\
\textbf{Lemma 3.7:} We have $\frac{\hat{\mathbb{E}}_1\square\
\hat{\mathbb{E}}_2[|B_t|^3]}{t}\rightarrow\ \ 0,$ as $ t\rightarrow\
0.$
\\
Proof:  Since $\varphi(x)=|x|^3 $ is convex, we obtain due to Lemma
3.6 that:
$$
\hat{\mathbb{E}}_1\square\
\hat{\mathbb{E}}_2[|B_t|^3]=\hat{\mathbb{E}}_3[|B_t|^3]=\overline{\sigma}_1^3\mathbb{E}[|W_1|^3]t^{3/2},$$
 where $(W_t)_{t\geq0}$ is Brownian motion under the linear expectation
 $\mathbb{E}$. The statement follows now easily.\\\\
\textbf{Proposition 3.8:} We have $$\hat{\mathbb{E}}_1\square\
\hat{\mathbb{E}}_2[\varphi(B_t-B_s)]=\hat{\mathbb{E}}_1\square\
\hat{\mathbb{E}}_2[\varphi(B_{t-s})],\ \ t\geq s\geq 0,\varphi\in
C_{l,lip}(\mathbb{R}).$$ The proof of Proposition 3.8 is rather
technical. To improve the readability of the paper, the proof is
postponed to the annex.\\\\
\textbf{Lemma 3.9:} For each $t\geq s$, $B_t-B_s$ is independent of
$(B_{t_1},B_{t_2},...,B_{t_n})$ under the sub-linear expectation
$\hat{\mathbb{E}}_1\square\ \hat{\mathbb{E}}_2[\cdot]$, for each
$n\in\mathbb{N},0\leq t_1,...,t_n \leq s$, that is, for all
$\varphi\in C_{l,lip}(\mathbb{R}^{n+1})$
\begin{eqnarray*}
&&\ \ \ \ \hat{\mathbb{E}}_1\square\
\hat{\mathbb{E}}_2[\varphi(B_{t_1},B_{t_2},...,B_{t_n},B_t-B_s)]\\
&&=\hat{\mathbb{E}}_1\square\
\hat{\mathbb{E}}_2[\hat{\mathbb{E}}_1\square\
\hat{\mathbb{E}}_2[\varphi(x_1,...,x_n,B_t-B_s)]|_{(x_1,...,x_n)=(B_{t_1},...,B_{t_n})}].
\end{eqnarray*}
We shift also the proof of Lemma 3.9 to the annex.\\\\
We are now able to give the proof of Theorem 3.2:\\\\
 \textbf{Proof (of Theorem 3.2):} It is sufficient to apply Lemma 3.4.
Due to the above statements, we know that the canonical process
$(B_t)_{t\geq0}$ is a G-Brownian motion under the sublinear
expectation
 $\hat{\mathbb{E}}_1\square\hat{\mathbb{E}}_2[\cdot].$ Consequently  $\hat{\mathbb{E}}_1\square\hat{\mathbb{E}}_2[\cdot]$  is a
 G-expectation on the space $(\Omega,\mathcal {H})$ and has the
 driver $G_1\Box G_2=G_{\underline{\sigma}_2,\overline{\sigma}_1}.$ \ \ \
 \ \ \ \ \ \ \ \ \ \ \ \ \ \  \ \ \ \ \ \ \ \ \ \ \ \ \ \ \ \ \ \ \ \ \ \
 \ \ \ \ \ \ \ \ \ \ \ \ \ \ \ \ \ \ \ \ \ \ \ \ \ \ \ \ \ \ \ \ \ \ \ \ \ \ \ \ \ \
  \ \ \ \ \ \ \ \ \ \ \ \ \ \ \ \ \ $\blacksquare$\\\\
Given n sublinear expectations
$\hat{\mathbb{E}}_1,...,\hat{\mathbb{E}}_n$ we define iteratively
$$\hat{\mathbb{E}}_1\square\hat{\mathbb{E}}_2\square\hat{\mathbb{E}}_3:=(\hat{\mathbb{E}}_1\square\hat{\mathbb{E}}_2)\square\hat{\mathbb{E}}_3,$$
and
$$\hat{\mathbb{E}}_1\square\hat{\mathbb{E}}_2\square...\square\hat{\mathbb{E}}_k:
=(\hat{\mathbb{E}}_1\square\hat{\mathbb{E}}_2\square...\square\hat{\mathbb{E}}_{k-1})\square\hat{\mathbb{E}}_k,\
3\leq k\leq n.$$\\
Then from Theorem 3.2 it follows:\\ \\
\textbf{Corollary 3.10:} Let  $0\leq
\underline{\sigma}_i\leq\overline{\sigma}_i,\ 1\leq i\leq n,$ and
denote by $\hat{\mathbb{E}}_i[\cdot]$ the $G_{\underline{\sigma}_i,\
\overline{\sigma}_i}$-expectation on the space $(\Omega,\mathcal
{H}).$ Then under the assumption\\
$\bigcap_{i=1}^n[\underline{\sigma}_i,\overline{\sigma}_i]\neq
\emptyset,$
$\hat{\mathbb{E}}_1\square\hat{\mathbb{E}}_2\square...\square\hat{\mathbb{E}}_n[\cdot]$
also is a G-expectation and has the driver
$G_{\underline{\sigma}_1,\overline{\sigma}_1}\square
G_{\underline{\sigma}_2,\overline{\sigma}_2} \square...\square
G_{\underline{\sigma}_n,\overline{\sigma}_n}.$ Moreover, for any
permutation $i_1,...,i_n$ of the natural numbers 1,...,n it holds:
$$\hat{\mathbb{E}}_1\square\hat{\mathbb{E}}_2\square...\square\hat{\mathbb{E}}_n[\cdot]=\hat{\mathbb{E}}_{i_1}\square\hat{\mathbb{E}}_{i_2}\square...\square
\hat{\mathbb{E}}_{i_n}[\cdot].$$ \\ \\
\textbf{Remark:} If
$\bigcap_{i=1}^n[\underline{\sigma}_i,\overline{\sigma}_i]$ is
empty, then
$\hat{\mathbb{E}}_1\square\hat{\mathbb{E}}_2\square...\square\hat{\mathbb{E}}_n[\cdot]=-\infty$,
otherwise
$\hat{\mathbb{E}}_1\square\hat{\mathbb{E}}_2\square...\square\hat{\mathbb{E}}_n[\cdot]$
is a $G_{\underline{\sigma},\overline{\sigma}}$-expectation, where
$[\underline{\sigma},\overline{\sigma}]=\bigcap_{i=1}^n[\underline{\sigma}_i,\overline{\sigma}_i]$.\\\\

\section{Annex}

\subsection{Proof of Proposition 3.8}We begin with the
proof of Proposition 3.8. For this we need the following two lemmas.
\\\\
\textbf{Lemma 4.1:} For all $T>0$ and all $X\in \mathcal {H}_T,$ we
have
$$\inf_{F\in \mathcal
{H}_T}\{\hat{\mathbb{E}}_1[X-F]+\hat{\mathbb{E}}_2[F]\}=\inf_{F\in
\mathcal {H}}\{\hat{\mathbb{E}}_1[X-F]+\hat{\mathbb{E}}_2[F]\}.$$
Proof:  From $\mathcal {H}_T\subseteq \mathcal {H}$ we see that
$$\inf_{F\in \mathcal
{H}_T}\{\hat{\mathbb{E}}_1[X-F]+\hat{\mathbb{E}}_2[F]\}\geq\inf_{F\in
\mathcal {H}}\{\hat{\mathbb{E}}_1[X-F]+\hat{\mathbb{E}}_2[F]\}.$$
Thus it remains to prove  the converse inequality.
\\
First we notice that, due to Proposition 2.8 and the subadditivity
of $\hat{\mathbb{E}}_3$, for any $F \in \mathcal {H}$,
$$\hat{\mathbb{E}}_2[F|\mathcal {H}_T]+
\hat{\mathbb{E}}_1[-F|\mathcal
{H}_T]\geq\hat{\mathbb{E}}_3[F|\mathcal {H}_T]+
\hat{\mathbb{E}}_3[-F|\mathcal {H}_T]\geq 0.$$ Consequently, for all
$X\in \mathcal {H}_T$ and all $F\in\mathcal {H},$
\begin{eqnarray*}
&&\ \ \ \ \hat{\mathbb{E}}_1[X-F]+\hat{\mathbb{E}}_2[F]\\
&&=\hat{\mathbb{E}}_1[\hat{\mathbb{E}}_1[X-F|\mathcal
{H}_T]]+\hat{\mathbb{E}}_2[F]\\
&&=\hat{\mathbb{E}}_1[X+\hat{\mathbb{E}}_1[-F|\mathcal
{H}_T]]+\hat{\mathbb{E}}_2[F]\\
&&=\hat{\mathbb{E}}_1[X-(-\hat{\mathbb{E}}_1[-F|\mathcal
{H}_T])]+\hat{\mathbb{E}}_2[-\hat{\mathbb{E}}_1[-F|\mathcal
{H}_T]]\\
&&\ \ \ \ -\hat{\mathbb{E}}_2[-\hat{\mathbb{E}}_1[-F|\mathcal
{H}_T]]+\hat{\mathbb{E}}_2[\hat{\mathbb{E}}_2[F|\mathcal {H}_T]]\\
&&\geq\hat{\mathbb{E}}_1[X-(-\hat{\mathbb{E}}_1[-F|\mathcal
{H}_T])]+\hat{\mathbb{E}}_2[-\hat{\mathbb{E}}_1[-F|\mathcal
{H}_T]]\\
&&\geq\inf_{F\in \mathcal
{H}_T}\{\hat{\mathbb{E}}_1[X-F]+\hat{\mathbb{E}}_2[F]\}.
\end{eqnarray*}
The statement now follows easily.\ \ \ \ \ \ \ \ \ \ \
\ \ \ \ \ \ \ \ \ \ \ \ \ \ \ \ \ \ \ \ \ \ \ \ \ \ \ \  \ \ \ \ \ \ \ \ \ \ \ \ \ \ $\blacksquare$\\\\
\textbf{Lemma 4.2:} For all $X \in \mathcal {H}_t^s,0\leq s\leq t,$
the following holds true:$$\inf_{F\in \mathcal
{H}_t}\{\hat{\mathbb{E}}_1[X-F]+\hat{\mathbb{E}}_2[F]\}=\inf_{F\in
\mathcal
{H}_t^s}\{\hat{\mathbb{E}}_1[X-F]+\hat{\mathbb{E}}_2[F]\}.$$ Proof:
Firstly, from $\mathcal {H}_t^s\subseteq\mathcal {H}_t$, we have,
obviously, for all $X \in \mathcal {H}_t^s,$
$$\inf_{F\in \mathcal
{H}_t}\{\hat{\mathbb{E}}_1[X-F]+\hat{\mathbb{E}}_2[F]\}\leq\inf_{F\in
\mathcal
{H}_t^s}\{\hat{\mathbb{E}}_1[X-F]+\hat{\mathbb{E}}_2[F]\}.$$
Secondly, for any $X \in \mathcal {H}_t^s$ and $ F \in\mathcal
{H}_t$, we can suppose without loss of generality that
$X=\varphi(B_{t_1}-B_s,...,B_{t_n}-B_{s})$ and
$F=\psi(B_{t'_1},B_{t'_2},...,B_{t'_k},B_{t_1}-B_{s},...,B_{t_n}-B_{s}),$
where $t'_1,...,t'_k\in[0,s],\ t_1,...,t_n\in[s,t], n,k \in
\mathbb{N}, \varphi\in C_{l,lip}(\mathbb{R}^{n})$ and $\psi\in
C_{l,lip}(\mathbb{R}^{n+k})$.\\
To simplify the notation we put:
$$Y_1=(B_{t'_1},B_{t'_2},...,B_{t'_k}),Y_2=(B_{t_1}-B_{s},...,B_{t_n}-B_{s}),\mathbf{x}=(x_1,x_2,...,x_k).$$
Then,
\begin{eqnarray*}
&&\ \ \ \ \hat{\mathbb{E}}_1[X-F]+\hat{\mathbb{E}}_2[F]\\
&&=\hat{\mathbb{E}}_1[\hat{\mathbb{E}}_1[\varphi(Y_2) -\psi(Y_1,Y_2)|\mathcal {H}_s]]+\hat{\mathbb{E}}_2[F]\\
&&=\hat{\mathbb{E}}_1[\hat{\mathbb{E}}_1[\varphi(Y_2)-\psi(\mathbf{x},Y_2)]|_{\mathbf{x}=Y_1}]+\hat{\mathbb{E}}_2[F]\\
&&=\hat{\mathbb{E}}_1[(\hat{\mathbb{E}}_1[\varphi(Y_2)-\psi(\mathbf{x},Y_2)]+\hat{\mathbb{E}}_2[\psi(\mathbf{x},Y_2)]-\hat{\mathbb{E}}_2[\psi(\mathbf{x},Y_2)])|_{\mathbf{x}=Y_1}]+\hat{\mathbb{E}}_2[F]\\
&&\geq\hat{\mathbb{E}}_1[\inf_{F\in \mathcal
{H}_t^s}\{\hat{\mathbb{E}}_1[X-F]+\hat{\mathbb{E}}_2[F]\}-\hat{\mathbb{E}}_2[\psi(\mathbf{x},Y_2)]|_{\mathbf{x}=Y_1}]+\hat{\mathbb{E}}_2[F]\\
&&=\inf_{F\in \mathcal {H}_t^s}\{\hat{\mathbb{E}}_1[X-F]
+\hat{\mathbb{E}}_2[F]\}+\hat{\mathbb{E}}_1[-\hat{\mathbb{E}}_2[\psi(\mathbf{x},Y_2)]|_{\mathbf{x}=Y_1}]\\
&&\ \ \ +\hat{\mathbb{E}}_2[\hat{\mathbb{E}}_2[\psi(\mathbf{x},Y_2)]|_{\mathbf{x}=Y_1}]\\
&&\geq\inf_{F\in \mathcal {H}_t^s}\{\hat{\mathbb{E}}_1[X-F]
+\hat{\mathbb{E}}_2[F]\}.
\end{eqnarray*}
Thus the proof is complete now.\ \ \ \ \ \ \ \ \ \ \ \ \ \ \ \ \ \ \ \ \ \
\ \ \ \ \ \ \ \ \ \ \ \ \ \ \ \ \ \ \ \ \ \ \ \ \ \ \ \ \ \ \ \ \ \   $\blacksquare$ \\\\
Now we are able to prove Proposition 3.8.\\
 \\
\textbf{ Proof (of Proposition 3.8)}: For arbitrarily fixed
$s\geq0$, we put $\widetilde{B}_{t}=B_{t+s}-B_s,\ t\geq0.$
 Then, obviously, $\mathcal {H}_{t+s}^s=\widetilde{\mathcal
{H}}_{t}$, $t\geq0,$ where $\widetilde{\mathcal {H}}_{t}$ is
generated by $\widetilde{B}_{t}$. Moreover, $\widetilde{B}_{t}$ is a
G-Brownian
Motion under $\hat{\mathbb{E}}_1$ and $\hat{\mathbb{E}}_2.$ \\
According to the Lemmas 4.1 and 4.2, we have the following:
\begin{eqnarray*}
&&\ \ \ \ \hat{\mathbb{E}}_1\square\
\hat{\mathbb{E}}_2[\varphi(B_t-B_s)]\\
&&=\inf_{F\in \mathcal
{H}_t^s}\{\hat{\mathbb{E}}_1[\varphi(B_t-B_s)-F]+\hat{\mathbb{E}}_2[F]\}\\
&&=\inf_{F\in \widetilde{\mathcal
{H}}_{t-s}}\{\hat{\mathbb{E}}_1[\varphi(\tilde{B}
_{t-s})-F]+\hat{\mathbb{E}}_2[F]\}\\
&&=\inf_{F\in {\mathcal {H}}_{t-s}}\{\hat{\mathbb{E}}_1[\varphi({B}
_{t-s})-F]+\hat{\mathbb{E}}_2[F]\}\\
 &&=\hat{\mathbb{E}}_1\square\
\hat{\mathbb{E}}_2[\varphi(B_{t-s})].
\end{eqnarray*}
Thus the proof of Proposition 3.8 is complete now.\ \ \ \ \ \ \ \ \ \ \ \ \ \ \ \ \ \ \ \ \ \ \ \ \ \ \ \ \ \ \ \ \ \ $\blacksquare$\\\\
\subsection{ Proof of Lemma 3.9}
Let us come now to the proof of Lemma 3.9, which we split into a sequel of lemmas.\\\\
 \textbf{Lemma 4.3:} For all $\varphi\in C_{l,lip}(\mathbb{R}^{n+1}),n\in \mathbb{N}$ and $0\leq
t_1,...,t_n \leq s\leq t,$ it holds:
\begin{eqnarray*}
&&\ \ \ \  \hat{\mathbb{E}}_1\square\
\hat{\mathbb{E}}_2[\varphi(B_{t_1},B_{t_2},...,B_{t_n},B_t-B_s)]\\
&&\geq \hat{\mathbb{E}}_1\square\
\hat{\mathbb{E}}_2[\hat{\mathbb{E}}_1\square\
\hat{\mathbb{E}}_2[\varphi(x_1,...,x_n,B_t-B_s)]|_{(x_1,...,x_n)=(B_{t_1},...,B_{t_n})}].
\end{eqnarray*}
Proof: Let $X=\varphi(B_{t_1},B_{t_2},...,B_{t_n},B_t-B_s).$ Without
loss of generality we can suppose
 that
$F\in \mathcal {H}$ has the form
$\psi(B_{t'_1},B_{t'_2},...,B_{t'_k},B_{t'_{k+1}}-B_{s},...,B_{t'_m}-B_{s}),$
where $0\leq t_1,...,t_n,t'_1,...,t'_k \leq s,\
t'_{k+1},...,t'_m\geq s ,m\geq k,m,k\in \mathbb{N},$ and $\varphi\in
C_{l,lip}(\mathbb{R}^{n+1}),\psi\in
C_{l,lip}(\mathbb{R}^{m}).$ \\    \\
For simplifying the notation we put: \begin{eqnarray*}
&&\mathbf{x}_1=(x_1,...,x_n),\mathbf{x}_2=(x'_1,...,x'_k),Y_1=(B_{t_1},B_{t_2},...,B_{t_n}),\\
&&Y_2=(B_{t'_1},...,B_{t'_k}),Y_3=(B_{t'_{k+1}}-B_{s},...,B_{t'_m}-B_{s}).
\end{eqnarray*}
 Then
\begin{eqnarray*}
&&\ \ \ \
\hat{\mathbb{E}}_1[X-F]+\hat{\mathbb{E}}_2[F]=\hat{\mathbb{E}}_1[\hat{\mathbb{E}}_1[X-F|\mathcal
{H}_s]]+\hat{\mathbb{E}}_2[\hat{\mathbb{E}}_2[F|\mathcal
{H}_s]]\\
&&=\hat{\mathbb{E}}_1[\hat{\mathbb{E}}_1[\varphi(\mathbf{x}_1,B_t-B_s)-\psi(\mathbf{x}_2,Y_3)]|_{\mathbf{x}_1=Y_1,\mathbf{x}_2=Y_2}]\\
&&\ \ \ \ +\hat{\mathbb{E}}_2[\hat{\mathbb{E}}_2[\psi(\mathbf{x}_2,Y_3)]|_{\mathbf{x}_2=Y_2}]\\
&&=\hat{\mathbb{E}}_1[(\hat{\mathbb{E}}_1[\varphi(\mathbf{x}_1,B_t-B_s)-\psi(\mathbf{x}_2,Y_3)]+\hat{\mathbb{E}}_2[\psi(\mathbf{x}_2,Y_3)]\\
&&\ \ \ -\hat{\mathbb{E}}_2[\psi(\mathbf{x}_2,Y_3)])|_{\mathbf{x}_1=Y_1,\mathbf{x}_2=Y_2}]+\hat{\mathbb{E}}_2[\hat{\mathbb{E}}_2[\psi(\mathbf{x}_2,Y_3)]|_{\mathbf{x}_2=Y_2}]\\
&&\geq\hat{\mathbb{E}}_1[(\hat{\mathbb{E}}_1\square\
\hat{\mathbb{E}}_2[\varphi(\mathbf{x}_1,B_t-B_s)]-\hat{\mathbb{E}}_2[\psi(\mathbf{x}_2,Y_3)])|_{\mathbf{x}_1=Y_1,\mathbf{x}_2=Y_2}]\\
&&\ \ \ \ +\hat{\mathbb{E}}_2[\hat{\mathbb{E}}_2[\psi(\mathbf{x}_2,Y_3)]|_{\mathbf{x}_2=Y_2}]\\
&&\geq\hat{\mathbb{E}}_1\square\
\hat{\mathbb{E}}_2[\hat{\mathbb{E}}_1\square\
\hat{\mathbb{E}}_2[\varphi(\mathbf{x}_1,B_t-B_s)]|_{\mathbf{x}_1=Y_1}]\\
&&=\hat{\mathbb{E}}_1\square\
\hat{\mathbb{E}}_2[\hat{\mathbb{E}}_1\square\
\hat{\mathbb{E}}_2[\varphi(x_1,...,x_n,B_t-B_s)]|_{(x_1,...,x_n)=(B_{t_1},...,B_{t_n})}].
\end{eqnarray*}
Hence, we get
\begin{eqnarray*}
&&\ \ \ \ \hat{\mathbb{E}}_1\square\
\hat{\mathbb{E}}_2[\varphi(B_{t_1},B_{t_2},...,B_{t_n},B_t-B_s)]\\
&&\geq\hat{\mathbb{E}}_1\square\
\hat{\mathbb{E}}_2[\hat{\mathbb{E}}_1\square\
\hat{\mathbb{E}}_2[\varphi(x_1,...,x_n,B_t-B_s)]|_{(x_1,...,x_n)=(B_{t_1},...,B_{t_n})}].
\end{eqnarray*}
The proof of the Lemma 4.3 is complete now.\ \ \ \ \ \ \ \ \ \ \ \ \ \
\ \ \ \ \ \ \ \ \ \ \ \ \ \ \ \ \ \ \ \ \ \ \ \ \ \ \ $\blacksquare$\\     \\
Let $Lip(\mathbb{R}^n),n\in\mathbb{N},$ denote the space of bounded
Lipschitz functions $\varphi\in Lip(\mathbb{R}^n)$ satisfying:
$$|\varphi(x)-\varphi(y)|\leq C|x-y| \ \ \ \ \ x,y\in \mathbb{R}^n,$$
where C is a constant only depending on $\varphi$.\\  \\
The proof that
\begin{eqnarray*} &&\ \ \ \ \hat{\mathbb{E}}_1\square\
\hat{\mathbb{E}}_2[\varphi(B_{t_1},B_{t_2},...,B_{t_n},B_t-B_s)]\\
&&\leq\hat{\mathbb{E}}_1\square\
\hat{\mathbb{E}}_2[\hat{\mathbb{E}}_1\square\
\hat{\mathbb{E}}_2[\varphi(x_1,...,x_n,B_t-B_s)]|_{(x_1,...,x_n)=(B_{t_1},...,B_{t_n})}]
\end{eqnarray*}
is much more difficult than that of the converse inequality. For the
proof we need the following
statements.\\\\
\textbf{Lemma 4.4}: We assume that the random variable
$\varphi(B_{t_1},B_{t_2}-B_{t_1},...,B_{t_n}-B_{t_{n-1}})$, with $
t_i\leq t_{i+1},\  i=1,...,n-1,\ n\in\mathbb{N} $ and $\varphi\in
Lip(\mathbb{R}^n),$ satisfies the following assumption: there exist
$L,M \geq 0$ s.t. $|\varphi|\leq L$, and $\varphi(x,y)=0$,  for all
 $(x,y)\in [-M,M]^c\times \mathbb{R}^{n-1}.$ \\
We define
\begin{eqnarray*}
&&\phi(x):\ =\hat{\mathbb{E}}_1\square \hat{\mathbb{E}}_2[\varphi(x,B_{t_2}-B_{t_1},...,B_{t_n}-B_{t_{n-1}})]\\
&&\ \ \ \ \ \ \ \ \ =\inf_{F\in \mathcal {H}_{t_n}^{t_1}}
\{\hat{\mathbb{E}}_1[\varphi(x,B_{t_2}-B_{t_1},...,B_{t_n}-B_{t_{n-1}})-F]+\hat{\mathbb{E}}_2[F]\}.
\end{eqnarray*}
Then we have the existence of an $\varepsilon$-optimal
$\widetilde{\psi}(x)$ of the form
$\psi(x,B_{t'_2}-B_{t_1},...,B_{t'_{l+1}}-B_{t_1})$, i.e., for any
$\varepsilon>0$  we can find a finite dimensional function
$\psi(x,\cdot)\in C_{l,lip}(\mathbb{R}^{l}),l\geq1,$ such that, for
suitable $ t'_2,....,t'_{l+1}\geq t_1$,

\begin{eqnarray*}
&&\widetilde{\psi}(x):=\hat{\mathbb{E}}_1[\varphi(x,B_{t_2}-B_{t_1},...,B_{t_n}-B_{t_{n-1}})\\
&&\ \ \ \ \ \ \ \  -\psi(x,B_{t'_2}-B_{t_1},...,B_{t'_{l+1}}-B_{t_1})]\\
&&\ \ \ \ \ \ \
+\hat{\mathbb{E}}_2[\psi(x,B_{t'_2}-B_{t_1},...,B_{t'_{l+1}}-B_{t_1})]
\end{eqnarray*}
satisfies $$|\widetilde{\psi}(x)-\phi(x)|\leq\varepsilon.$$ \\
Proof: Since $\varphi \in Lip(\mathbb{R}^n)$, we find for any $
\varepsilon>0$ some sufficiently large  $  J \geq 1$ s.t. for all
$x,\tilde{x}\in \mathbb{R}$ with $
|x-\widetilde{x}|\leq\frac{2M}{J}$ it holds
$|\varphi(x,y)-\varphi(\widetilde{x},y)|\leq \varepsilon/6.$ We then
let $-M=x_0\leq x_1\leq....\leq x_J=M ,$ be such that
$|x_{j+1}-x_j|=\frac{2M}{J}, \ 0\leq j\leq J-1.$
\\
On the other hand, for every fixed $j$ there are some $m_j\geq1$, $
 t_{i,j}\geq t_1$ $(2\leq i\leq m_j)$ and $\psi^{x_j}\in
C_{l,lip}(\mathbb{R}^{m_j-1}),$ such that
\begin{eqnarray*}
&&\phi(x_j)\leq \hat{\mathbb{E}}_1[\varphi(x_j,B_{t_2}-B_{t_1},...,B_{t_n}-B_{t_{n-1}})\\
&&\ \ \ \ \ \ \ \ \ \
-\psi^{x_j}(B_{t_{2,j}}-B_{t_1},...,B_{t_{m_j,j}}-B_{t_1})]\\
&&\ \ \ \ \ \ \ \
+\hat{\mathbb{E}}_2[\psi^{x_j}(B_{t_{2,j}}-B_{t_1},...,B_{t_{m_j,j}}-B_{t_1})]\\
&&\ \ \ \ \ \ \ \  \leq\phi(x_j) +\varepsilon/6.
\end{eqnarray*}
Since there are only a finite number of $j$ we can find a finite
dimensional function denoted by $\psi(x_j,y),\ y\in \mathbb{R}^l,$
s.t. for each fixed j, $ \psi(x_j,\cdot) \in
C_{l,lip}(\mathbb{R}^l)$ and
$$\psi(x_j,B_{t'_2}-B_{t_1},...,B_{t'_{l+1}}-B_{t_1})=\psi^{x_j}(B_{t_{2,j}}-B_{t_1},...,B_{t_{m_j,j}}-B_{t_1}),$$
 where  $\{t'_2,..., t'_{l+1}\}=\bigcup_{j=1}^J\{t_{2,j},...,t_{m_j,j}\}$.\\  \\
With the convention $\psi(x_0,y)=\psi(x_J,y)=0,y\in \mathbb{R}^l,$
we define
\[
\psi(x,y):=\left\{\begin{tabular}{ll}$\frac{x_{j+1}-x}{x_{j+1}-x_j}\psi(x_j,y)+\frac{x-x_j}{x_{j+1}-x_j}\psi(x_{j+1},y),$&$x\in[x_j,x_{j+1}],$\\$ $&$0\leq j\leq J-1,$\\
0,&$\mbox{otherwise}.$
\end{tabular}\right.\]
Obviously, $\psi(x,y) \in C_{l,lip}(\mathbb{R}^{l+1})$. \\\\We now
introduce $\widetilde{\psi}(x):$
\begin{eqnarray*}
&&=\hat{\mathbb{E}}_1[\varphi(x,B_{t_2}-B_{t_1},...,B_{t_n}-B_{t_{n-1}})-\psi(x,B_{t'_2}-B_{t_1},...,B_{t'_{l+1}}-B_{t_1})]\\
&&\ \ \ \ \ \ \ \ \ \ \ \ \ \ \
+\hat{\mathbb{E}}_2[\psi(x,B_{t'_2}-B_{t_1},...,B_{t'_{l+1}}-B_{t_1})].
\end{eqnarray*}
If $x\notin[-M,M], $ $\varphi(x,\cdot)=0$ and $\psi(x,\cdot)=0.$
Consequently, $\widetilde{\psi}(x)=0.$ Moreover, from Proposition
3.5 we have that for $x\notin[-M,M]$ also $\phi(x)=0.$
 Then  $\widetilde{\psi}(x)= \phi(x)=0$ when $x\notin[-M,M], $ and
 we have also
$|\widetilde{\psi}(x_j)- \phi(x_j)|\leq\varepsilon/6$ for each j. We
also recall that, for all $0\leq j\leq J-1$ and all $x\in[x_j,
x_{j+1}],$
$$|\varphi(x,y)-\varphi(x_j,y)|\leq\varepsilon/6, \ \mbox{for all}\
y\in\mathbb{R}^{n-1}.$$ Our objective is to estimate
$$|\widetilde{\psi}(x)- \phi(x)|\leq
|\widetilde{\psi}(x)-\phi(x_j)|+|\phi(x_j)-\phi(x)|.$$ For this end
we notice that, with the notation:
$$Y_1=(B_{t_2}-B_{t_1},...,B_{t_n}-B_{t_{n-1}}),Y_2=(B_{t'_2}-B_{t_1},...,B_{t'_{l+1}}-B_{t_{1}}),$$
we have from the definition of $\phi(x)$ and $\phi(x_j)$ and from
the properties of $\hat{\mathbb{E}}_1\square \hat{\mathbb{E}}_2$ as
sublinear expectation:
$$|\phi(x)-\phi(x_j)|\leq \hat{\mathbb{E}}_1\square
\hat{\mathbb{E}}_2[|\varphi(x,Y_1)-\varphi(x_j,Y_1)|]\leq\varepsilon/6.$$
On the other hand, since
$|\varphi(x,Y_1)-\varphi(x_j,Y_1)|\leq\varepsilon/6,$
\begin{eqnarray*}
&&\ \ \ \ |\widetilde{\psi}(x)-\phi(x_j)|\\
&&=|\hat{\mathbb{E}}_1[\varphi(x,Y_1)-\psi(x,Y_2)]+\hat{\mathbb{E}}_2[\psi(x,Y_2)]-\phi(x_j)|\\
&&
 \leq\mid\hat{\mathbb{E}}_1[\varphi(x_j,Y_1)-\psi(x,Y_2)]+\hat{\mathbb{E}}_2[\psi(x,Y_2)]-\phi(x_j)\mid+\varepsilon/6.
\end{eqnarray*}
Due to the definition of $\phi(x_j)$, the latter expression without
module is non-negative. Thus,
 \begin{eqnarray*}
 &&\ \ \ \ |\widetilde{\psi}(x)-\phi(x_j)|\\
&&\leq\hat{\mathbb{E}}_1[\varphi(x_j,Y_1)-\psi(x,Y_2)]+\hat{\mathbb{E}}_2[\psi(x,Y_2)]-\phi(x_j)+\varepsilon/6\\
&&\leq\hat{\mathbb{E}}_1[\frac{x_{j+1}-x}{x_{j+1}-x_j}(\varphi(x_j,Y_1)-\psi(x_j,Y_2))+\frac{x-x_j}{x_{j+1}-x_j}(\varphi(x_{j+1},Y_1)\\
&&  -\psi(x_{j+1},Y_2))]
+\hat{\mathbb{E}}_2[\frac{x_{j+1}-x}{x_{j+1}-x_j}\psi(x_j,Y_2)+\frac{x-x_j}{x_{j+1}-x_j}\psi(x_{j+1},Y_2)]\\
&& -\phi(x_j)+2\varepsilon/6\\
&&\leq\frac{x_{j+1}-x}{x_{j+1}-x_j}\{\hat{\mathbb{E}}_1[\varphi(x_j,Y_1)-\psi(x_j,Y_2)]+\hat{\mathbb{E}}_2[\psi(x_j,Y_2)]-\phi(x_j)\}\\
&&+\frac{x-x_j}{x_{j+1}-x_j}\{\hat{\mathbb{E}}_1[\varphi(x_{j+1},Y_1)-\psi(x_{j+1},Y_2)]+\hat{\mathbb{E}}_2[\psi(x_{j+1},Y_2)]-\phi(x_j)\}\\
&&+2\varepsilon/6.
\end{eqnarray*}
Hence, due to the choice of $\psi^{x_j}$ and $\psi^{x_{j+1}}$,
$$|\widetilde{\psi}(x)-\phi(x_j)|\leq5\varepsilon/6.$$
This latter estimate combined with the fact that for
$|\phi(x)-\phi(x_j)|\leq\varepsilon/6$ then yields
$$|\widetilde{\psi}(x)- \phi(x)|\leq \varepsilon.$$
The proof of Lemma 4.4 is complete now.\ \ \ \ \ \ \ \ \ \ \ \ \ \ \ \ \ \ \ \ \ \ \ \ \ \ \ \ \ \ \ \ \ \ \ \ \ \ \ \ \ \  $\blacksquare$ \\       \\
Lemma 4.4 allows to prove the following:\\\\
 \textbf{Lemma
4.5:} Let $\varphi\in \ Lip(\mathbb{R}^n)$ be bounded and such that,
for some real $M>0,$
$\mbox{supp}(\varphi)\subset[-M,M]\times\mathbb{R}^{n-1}.$ Then, for
all $0\leq t_1\leq t_2...\leq t_n$,
\begin{eqnarray*}
&&\ \ \ \ \hat{\mathbb{E}}_1\square\hat{\mathbb{E}}_2[\varphi(B_{t_1},B_{t_2}-B_{t_1},...,B_{t_n}-B_{t_{n-1}})]\\
&&=\hat{\mathbb{E}}_1\square\hat{\mathbb{E}}_2[\hat{\mathbb{E}}_1\square\hat{\mathbb{E}}_2[\varphi(x,B_{t_2}-B_{t_1},...,B_{t_n}-B_{t_{n-1}})]|_{x=B_{t_1}}]].
\end{eqnarray*}
Proof: Firstly, it follows directly from Lemma 4.3 that:
\begin{eqnarray*}
&&\ \ \ \ \hat{\mathbb{E}}_1\square\hat{\mathbb{E}}_2[\varphi(B_{t_1},B_{t_2}-B_{t_1},...,B_{t_n}-B_{t_{n-1}})]\\
&&\geq\hat{\mathbb{E}}_1\square\hat{\mathbb{E}}_2[\hat{\mathbb{E}}_1\square\hat{\mathbb{E}}_2[\varphi(x,B_{t_2}-B_{t_1},...,B_{t_n}-B_{t_{n-1}})]|_{x=B_{t_1}}]].\
\ \ \ (1)
\end{eqnarray*}
 Secondly, from Lemma 4.4 we know that for any $\varepsilon>0$
 there is some $\psi\in C_{l,lip}(\mathbb{R}^{l+1})$ such that
 $ |\widetilde{\psi}(x)-\phi(x)|\leq \varepsilon ,$ for all $x
\in\mathbb{R},$ where $\widetilde{\psi}(x)$ and $\phi(x)$ have been introduced in Lemma 4.4 .\\
Due to Lemma 4.1, there is
$\widetilde{\phi}(B_{t''_1},....,B_{t''_k})\in\mathcal
{H}_{t_1},0\leq t''_1,...,t''_k\leq t_1,k\in\mathbb{N}$, such that
$$|\hat{\mathbb{E}}_1[\phi(B_{t_1})-\widetilde{\phi}(B_{t''_1},....,B_{t''_k})]+\hat{\mathbb{E}}_2[\widetilde{\phi}(B_{t''_1},....,B_{t''_k})]
-\hat{\mathbb{E}}_1\square\hat{\mathbb{E}}_2[\phi(B_{t_1})]|\leq\varepsilon.$$
For $t'_2,...,t'_{l+1}\geq t_1$ from the definition of
$\widetilde{\psi}(x)$ in Lemma 4.4 we put $$\psi'(x)=
\hat{\mathbb{E}}_2[\psi(x,B_{t'_2}-B_{t_1},...,B_{t'_{l+1}}-B_{t_1})]$$
and $$ F=\psi(B_{t_1},B_{t'_2}-B_{t_1},...,B_{t'_{l+1}}-B_{t_1})
+\widetilde{\phi}(B_{t''_1},....,B_{t''_k})-\psi'(B_{t_1}).$$ Notice
that $$\hat{\mathbb{E}}_2[F|\mathcal
{H}_{t_1}]=\widetilde{\phi}(B_{t''_1},....,B_{t''_k})$$ and
$$\hat{\mathbb{E}}_1[\varphi(B_{t_1},B_{t_2}-B_{t_1},...,B_{t_n}-B_{t_{n-1}})-F|\mathcal
{H}_{t_1}]=\widetilde{\psi}(B_{t_1})-\widetilde{\phi}(B_{t''_1},....,B_{t''_k}).$$
Then, due to the choice of
$\widetilde{\phi}(B_{t''_1},....,B_{t''_k}),$
\begin{eqnarray*}
&&\ \ \ \ \hat{\mathbb{E}}_1\square\hat{\mathbb{E}}_2[\varphi(B_{t_1},B_{t_2}-B_{t_1},...,B_{t_n}-B_{t_{n-1}})]-\hat{\mathbb{E}}_1\square\hat{\mathbb{E}}_2[\phi(B_{t_1})]\\
&&\leq\hat{\mathbb{E}}_1[\varphi(B_{t_1},B_{t_2}-B_{t_1},...,B_{t_n}-B_{t_{n-1}})-F]+\hat{\mathbb{E}}_2[\hat{\mathbb{E}}_2[F|\mathcal
{H}_{t_1}]]\\
&&\ \ \ -(\hat{\mathbb{E}}_1[\phi(B_{t_1})-\widetilde{\phi}(B_{t''_1},....,B_{t''_k})]+\hat{\mathbb{E}}_2[\widetilde{\phi}(B_{t''_1},....,B_{t''_k})])+\varepsilon\\
&&=\hat{\mathbb{E}}_1[\varphi(B_{t_1},B_{t_2}-B_{t_1},...,B_{t_n}-B_{t_{n-1}})-F]\\
&&\ \ \ \ \ -\hat{\mathbb{E}}_1[\phi(B_{t_1})-\widetilde{\phi}(B_{t''_1},....,B_{t''_k})]+\varepsilon\\
&&=\hat{\mathbb{E}}_1[\hat{\mathbb{E}}_1[\varphi(B_{t_1},B_{t_2}-B_{t_1},...,B_{t_n}-B_{t_{n-1}})-F|\mathcal
{H}_{t_1}]]\\
&&\ \ \ -\hat{\mathbb{E}}_1[\phi(B_{t_1})-\widetilde{\phi}(B_{t''_1},....,B_{t''_k})]+\varepsilon\\
&&=\hat{\mathbb{E}}_1[\widetilde{\psi}(B_{t_1})-\widetilde{\phi}(B_{t''_1},....,B_{t''_k})]
-\hat{\mathbb{E}}_1[\phi(B_{t_1})-\widetilde{\phi}(B_{t''_1},....,B_{t''_k})]+\varepsilon\\
&&\leq\hat{\mathbb{E}}_1[|\phi(B_{t_1})-\widetilde{\psi}(B_{t_1})|]+\varepsilon\\
&&\leq 2\varepsilon.
\end{eqnarray*}
From the definition of $\phi$ in Lemma 4.4 and the arbitrariness of
$\varepsilon>0$ it follows then that
\begin{eqnarray*}
&&\ \ \ \ \hat{\mathbb{E}}_1\square\hat{\mathbb{E}}_2[\varphi(B_{t_1},B_{t_2}-B_{t_1},...,B_{t_n}-B_{t_{n-1}})]\\
&&\leq\hat{\mathbb{E}}_1\square\hat{\mathbb{E}}_2[\hat{\mathbb{E}}_1\square\hat{\mathbb{E}}_2[\varphi(x,B_{t_2}-B_{t_1},...,B_{t_n}-B_{t_{n-1}})]|_{x=B_{t_1}}]].
\end{eqnarray*}
This together with $(1)$  yields the wished statement. The proof
of Lemma 4.5 is complete now.\ \ \ \ \ \ \ \ \ \ \ \ \ \ \ \ \ \ \ \ \ \ \ \
\ \ \ \ \ \ \ \ \ \ \ \ \ \ \ \ \ \ \ \ \ \ \ \ \ \ \ \ \ \ \ \ \ \ \ \ \ \ \ \ \ \ \ \ \ \ $\blacksquare$\\\\
In the next statement we extend Lemma 4.5 to general functions of
$Lip(\mathbb{R}^n).$\\\\
\textbf{Lemma 4.6: } Let $\varphi \in Lip(\mathbb{R}^n)$, $n\geq1,$
and $ t_n\geq t_{n-1}\geq...\geq t_1\geq0.$ Then
\begin{eqnarray*}
&&\ \ \ \ \hat{\mathbb{E}}_1\square\hat{\mathbb{E}}_2[\varphi(B_{t_1},B_{t_2}-B_{t_1},...,B_{t_n}-B_{t_{n-1}})]\\
&&=\hat{\mathbb{E}}_1\square\hat{\mathbb{E}}_2[\hat{\mathbb{E}}_1\square\hat{\mathbb{E}}_2[\varphi(x,B_{t_2}-B_{t_1},...,B_{t_n}-B_{t_{n-1}})]|_{x=B_{t_1}}]].
\end{eqnarray*}
Proof: Let $L>0$ be such that $|\varphi|\leq L.$ Given an
arbitrarily large $M>0$ we define, for all $y\in \mathbb{R}^{n-1},$

\[
\widetilde{\varphi}(x,y):=\left\{\begin{tabular}{ll}$\varphi(x,y),$\ \ \ &$x\in[-M,M]$\\
$\varphi(-M,y)(M+1+x),$\ \ \ \ &$x\in[-M-1,-M]$\\
$\varphi(M,y)(M+1-x),$\ \ \ \ \ \ &$x\in[M,M+1]$\\
$0,$\ \ \ \ \ \ \ &$\mbox{otherwise}.$ \end{tabular}\right.\]
Obviously,
 $\widetilde{\varphi}$ satisfies the
assumptions of Lemma 4.5. \\ Letting
$$\widetilde{\varphi}'(x)=\hat{\mathbb{E}}_1\square\hat{\mathbb{E}}_2[\widetilde{\varphi}(x,B_{t_2}-B_{t_1},...,B_{t_n}-B_{t_{n-1}})]$$
and
$$\phi(x)=\hat{\mathbb{E}}_1\square\hat{\mathbb{E}}_2[\varphi(x,B_{t_2}-B_{t_1},...,B_{t_n}-B_{t_{n-1}})],$$
we have
\begin{eqnarray*} &&\ \ \ \ |
\phi(x)-\widetilde{\varphi}'(x)|\\&&=|\hat{\mathbb{E}}_1\square\hat{\mathbb{E}}_2[\varphi(x,B_{t_2}-B_{t_1},...,B_{t_n}-B_{t_{n-1}})]\\
&&\ \  -\hat{\mathbb{E}}_1\square\hat{\mathbb{E}}_2[\widetilde{\varphi}(x,B_{t_2}-B_{t_1},...,B_{t_n}-B_{t_{n-1}})]|\\
&& \leq\hat{\mathbb{E}}_1\square\hat{\mathbb{E}}_2[|\varphi(x,B_{t_2}-B_{t_1},...,B_{t_n}-B_{t_{n-1}})\\
&&\ \ \  -\widetilde{\varphi}(x,B_{t_2}-B_{t_1},...,B_{t_n}-B_{t_{n-1}})|]\\
&& \leq\frac{2L}{M}|x|.
\end{eqnarray*}
Consequently,
\begin{eqnarray*}
&&|\hat{\mathbb{E}}_1\square\hat{\mathbb{E}}_2[\phi(B_{t_1})]-\hat{\mathbb{E}}_1\square\hat{\mathbb{E}}_2[\widetilde{\varphi}'(B_{t_1})]|
\leq
\hat{\mathbb{E}}_1\square\hat{\mathbb{E}}_2[|\phi(B_{t_1})-\widetilde{\varphi}'(B_{t_1})|]\\
&&\ \ \ \leq
\hat{\mathbb{E}}_1\square\hat{\mathbb{E}}_2[\frac{2L}{M}|B_{t_1}|]
=\frac{2L}{M}\hat{\mathbb{E}}_1\square\hat{\mathbb{E}}_2[|B_{t_1}|].
\end{eqnarray*}
On the other hand, from the definition of $\widetilde{\varphi}$ we
also obtain
\begin{eqnarray*}
&& \ \ \ \ \
|\hat{\mathbb{E}}_1\square\hat{\mathbb{E}}_2[\varphi(B_{t_1},B_{t_2}-B_{t_1},...,B_{t_n}-B_{t_{n-1}})]
\\&&\ \ \ \ \ \ \ \ \ -\hat{\mathbb{E}}_1\square\hat{\mathbb{E}}_2[\widetilde{\varphi}(B_{t_1},B_{t_2}-B_{t_1},...,B_{t_n}-B_{t_{n-1}})]
|\\
&&\leq\hat{\mathbb{E}}_1\square\hat{\mathbb{E}}_2[|\varphi(B_{t_1},B_{t_2}-B_{t_1},...,B_{t_n}-B_{t_{n-1}})\\
&&\ \ \ \ \ \ \ \ \ \ \ \ \ -\widetilde{\varphi}(B_{t_1},B_{t_2}-B_{t_1},...,B_{t_n}-B_{t_{n-1}})|]\\
&&\leq\frac{2L}{M}\hat{\mathbb{E}}_1\square\hat{\mathbb{E}}_2[|B_{t_1}|].
 \end{eqnarray*}
Thus, since due to Lemma 4.5
$$\hat{\mathbb{E}}_1\square\hat{\mathbb{E}}_2[\widetilde{\varphi}(B_{t_1},B_{t_2}-B_{t_1},...,B_{t_n}-B_{t_{n-1}})]
=\hat{\mathbb{E}}_1\square\hat{\mathbb{E}}_2[\widetilde{\varphi}'(B_{t_1})],$$
we get by letting $M\mapsto+\infty$ the relation
\begin{eqnarray*}
&&\ \ \ \ \ \hat{\mathbb{E}}_1\square\hat{\mathbb{E}}_2[\varphi(B_{t_1},B_{t_2}-B_{t_1},...,B_{t_n}-B_{t_{n-1}})]\\
&&=
\hat{\mathbb{E}}_1\square\hat{\mathbb{E}}_2[\hat{\mathbb{E}}_1\square\hat{\mathbb{E}}_2[\varphi(x,B_{t_2}-B_{t_1},...,B_{t_n}-B_{t_{n-1}})]|_{x=B_{t_1}}]].
\end{eqnarray*}
The proof of Lemma 4.6 is complete.\ \ \ \ \ \ \ \ \ \ \ \ \ \ \ \ \ \ \ \ \ \ \ \ \ \ \ \
\ \ \ \ \ \ \ \ \ \ \ \ \ \ \ \ \ \ \ \ \ \ \ \ \ \ \ \  \ \ \ \ \ \ \ \ \ \ \ \ \ \  $\blacksquare$\\\\
\textbf{Lemma 4.7:} For  all $\varphi\in Lip(\mathbb{R}^{n-1}),$ $n
\geq 1,$ and $0\leq t_1\leq t_2\leq...\leq t_n,$ we have
\begin{eqnarray*}
&&\ \ \ \hat{\mathbb{E}}_1\square\hat{\mathbb{E}}_2[\varphi(B_{t_2}-B_{t_1},...,B_{t_n}-B_{t_{n-1}})]\\
&&=
  \hat{\mathbb{E}}_1\square\hat{\mathbb{E}}_2[\hat{\mathbb{E}}_1\square\hat{\mathbb{E}}_2[
  \varphi(y,B_{t_3}-B_{t_2},...,B_{t_n}-B_{t_{n-1}})]|_{y=B_{t_2}-B_{t_1}}]
\end{eqnarray*}
Proof:  Lemma 4.2 allows to repeat the arguments of the Lemmas 4.3
to 4.6 in $\mathcal {H}_{t_1}^{t_n}.$  The result of Lemma 4.7 then
follows.\ \ \ \ \ \ \ \ \ \ \ \ \ \
\ \ \ \ \ \ \ \ \ \ \ \ \ \ \ \ \
\ \ \ \ \ \ \ \ \ \ \ $\blacksquare$
\\\\
Finally, we have:\\\\
 \textbf{Lemma 4.8:} Let $\varphi\in
Lip(\mathbb{R}^{n+1}),$ $n\geq1$ and
 $0\leq t_1,...,t_n \leq s.$ Then
\begin{eqnarray*}
&&\ \ \hat{\mathbb{E}}_1\square\hat{\mathbb{E}}_2[\varphi(B_{t_1},B_{t_2},...,B_{t_n},B_t-B_s)]\\
&&=\hat{\mathbb{E}}_1\square\hat{\mathbb{E}}_2[\hat{\mathbb{E}}_1\square\hat{\mathbb{E}}_2[\varphi(x_1,...,x_n,B_t-B_s)]|_{(x_1,...,x_n)=(B_{t_1},B_{t_2},...,B_{t_n})}]].
\end{eqnarray*}
Proof: Without any loss of generality we can suppose $0\leq t_1 \leq
t_2\leq...\leq t_n.$ Then there is some $\widetilde{\varphi}\in
Lip(\mathbb{R}^{n+1})$ such that
$\varphi(B_{t_1},B_{t_2},...,B_{t_n},B_t-B_s)=\widetilde{\varphi}(B_{t_1},B_{t_2}-B_{t_1},...,B_{t_n}-B_{t_{n-1}},B_t-B_s)\in\mathcal
{H}_t.$ With the notation $\mathbf{x}=(x_1,...,x_n),$ and due to the
Lemmas 4.1 to 4.7 we have
 \begin{eqnarray*}
 &&\ \ \hat{\mathbb{E}}_1\square\hat{\mathbb{E}}_2[\varphi(B_{t_1},B_{t_2},...,B_{t_n},B_t-B_s)]\\
 &&=\hat{\mathbb{E}}_1\square\hat{\mathbb{E}}_2[\widetilde{\varphi}(B_{t_1},B_{t_2}-B_{t_1},...,B_{t_n}-B_{t_{n-1}},B_t-B_s)]\\
 &&\ \ \ \ \ \ \ \ . \ \ \ .\ \ \ \ .\ \ \ \ \ .\ \ \\
 &&=\hat{\mathbb{E}}_1\square\hat{\mathbb{E}}_2[\hat{\mathbb{E}}_1\square\hat{\mathbb{E}}_2[\widetilde{\varphi}(\mathbf{x},B_t-B_s)]
 |_{\mathbf{x}=(B_{t_1},B_{t_2}-B_{t_1},...,B_{t_n}-B_{t_{n-1}})}]]
 \\
 &&=\hat{\mathbb{E}}_1\square\hat{\mathbb{E}}_2[\hat{\mathbb{E}}_1\square\hat{\mathbb{E}}_2[\varphi(\mathbf{x},B_t-B_s)]|_{\mathbf{x}=(B_{t_1},B_{t_2},...,B_{t_n})}]]\\
 &&=\hat{\mathbb{E}}_1\square\hat{\mathbb{E}}_2[\hat{\mathbb{E}}_1\square\hat{\mathbb{E}}_2[\varphi(x_1,...,x_n,B_t-B_s)]|_{(x_1,...,x_n)=(B_{t_1},B_{t_2},...,B_{t_n})}]].
  \end{eqnarray*}
The proof of Lemma 4.8 is complete now.\ \ \ \ \ \ \ \ \ \ \ \ \ \ \ \ \ \ \ \ \ \ \ \ \ \ \ \ \ \ \ \ \
 \ \ \ \ \ \ \ \ \ \ \ \ \ \ \ \ \ \ \ \ \ \ \  $\blacksquare$\\  \\
Let us now come to the proof of Lemma 3.9.\\
 \textbf{Proof (of Lemma 3.9)  :}
In a first step, we will prove that for each $\varphi \in
C_{l,lip}(\mathbb{R}^{n+1})$ there exists a sequence of bounded
Lipschitz functions $(\varphi_N)_{N\geq1}$ such that
\begin{eqnarray*}
&&\hat{\mathbb{E}}_1[|\varphi_N
(B_{t_1},B_{t_2},...,B_{t_n},B_t-B_s)-
\varphi(B_{t_1},B_{t_2},...,B_{t_n},B_t-B_s)|]\\
&&\longrightarrow 0,\ \ \mbox{as}\ \ N\longrightarrow\infty.
  \end{eqnarray*}
For this end we put
 $$l_N(x)=(x\wedge N)\vee(-N),N\geq1, \ x\in
\mathbb{R},$$ and  $$
\varphi_N(x_1,...,x_{n+1})=\varphi(l_N(x_1),...,l_N(x_{n+1})),$$ and
we notice that $$|x-l_N(x)|\leq\frac{|x|^2}{N},\ \mbox{for all}\
x\in\mathbb{R}.$$
Obviously, the functions $\varphi_N$ are bounded
and Lipschitz, and, moreover,
\begin{eqnarray*}
&&\ \ \ |\varphi_N(x_1,...,x_{n+1})-\varphi(x_1,...,x_{n+1})|\\
&&=|\varphi(l_N(x_1),...,l_N(x_{n+1}))-\varphi(x_1,...,x_{n+1})|\\
&&\leq C
(1+|x_1|^m+...+|x_{n+1}|^m)\sqrt{\sum_{i=1}^{n+1}\frac{|x_i|^4}{N^2}}\\
&&=\frac{C
(1+|x_1|^m+...+|x_{n+1}|^m)\sqrt{\sum_{i=1}^{n+1}|x_i|^4}}{N},
\end{eqnarray*}
where C and $m\geq0$ are constants only depending on $\varphi$.
Then, in virtue of the finiteness of
$\hat{\mathbb{E}}_1[(1+|B_{t_1}|^m+...+|B_{t_n}|^m+|B_t-B_s|^m)(\sum_{i=1}^{n}|B_{t_i}|^4+|B_t-B_s|^4)^{\frac{1}{2}}],$
we get\\\\ $\hat{\mathbb{E}}_1[|\varphi_N
(B_{t_1},B_{t_2},...,B_{t_n},B_t-B_s)-
\varphi(B_{t_1},B_{t_2},...,B_{t_n},B_t-B_s)|]\longrightarrow 0,$\\
as $N\longrightarrow\infty.$ \\\\Let $\mathbf{x}_1=(x_1,...,x_n)$
and $Y_1=(B_{t_1},B_{t_2},...,B_{t_n}).$ Then, due to our above
convergence result,
\begin{eqnarray*}
&&\ \ \
|\hat{\mathbb{E}}_1\square\hat{\mathbb{E}}_2[\varphi_N(Y_1,B_t-B_s)]
-\hat{\mathbb{E}}_1\square\hat{\mathbb{E}}_2[\varphi(Y_1,B_t-B_s)]|\\
&&\leq\hat{\mathbb{E}}_1\square\hat{\mathbb{E}}_2[|\varphi_N(Y_1,B_t-B_s)-\varphi(Y_1,B_t-B_s)|]\\
&&\leq
\hat{\mathbb{E}}_1[|\varphi_N(Y_1,B_t-B_s)-\varphi(Y_1,B_t-B_s)|]\\
&&\longrightarrow 0,\ \ \mbox{as}\ \ N\longrightarrow\infty,
\end{eqnarray*}
and, from Lemma 4.3,
\begin{eqnarray*}
&&\ \ \ \ \ |\hat{\mathbb{E}}_1\square\hat{\mathbb{E}}_2[\hat{\mathbb{E}}_1\square\hat{\mathbb{E}}_2[\varphi_N(\mathbf{x}_1,B_t-B_s)]|_{\mathbf{x}_1=Y_1}]]\\
&&\ \ \ \ \ -\hat{\mathbb{E}}_1\square\hat{\mathbb{E}}_2[\hat{\mathbb{E}}_1\square\hat{\mathbb{E}}_2[\varphi(\mathbf{x}_1,B_t-B_s)]|_{\mathbf{x}_1=Y_1}]]|\\
&&\leq\hat{\mathbb{E}}_1\square\hat{\mathbb{E}}_2[\hat{\mathbb{E}}_1\square\hat{\mathbb{E}}_2[|\varphi_N(\mathbf{x}_1,B_t-B_s)-\varphi(\mathbf{x}_1,
B_t-B_s)|]|_{\mathbf{x}_1=Y_1}]\\
&&\leq\hat{\mathbb{E}}_1[|\varphi_N(Y_1,B_t-B_s)-\varphi(Y_1,B_t-B_s)|]\\
&&\longrightarrow0,\ \ \mbox{as}\ \ N\longrightarrow\infty.
\end{eqnarray*}
On the other hand, from Lemma 4.8 we have
\begin{eqnarray*}
&&\ \ \hat{\mathbb{E}}_1\square\hat{\mathbb{E}}_2[\varphi_N(B_{t_1},B_{t_2},...,B_{t_n},B_t-B_s)]\\
&&=\hat{\mathbb{E}}_1\square\hat{\mathbb{E}}_2[\hat{\mathbb{E}}_1\square\hat{\mathbb{E}}_2[\varphi_N(x_1,...,x_n,B_t-B_s)]|_{(x_1,...,x_n)=(B_{t_1},B_{t_2},...,B_{t_n})}]].
\end{eqnarray*}
Combining the above results we can conclude that
\begin{eqnarray*}
&&\ \ \hat{\mathbb{E}}_1\square\hat{\mathbb{E}}_2[\varphi(B_{t_1},B_{t_2},...,B_{t_n},B_t-B_s)]\\
&&=\hat{\mathbb{E}}_1\square\hat{\mathbb{E}}_2[\hat{\mathbb{E}}_1\square\hat{\mathbb{E}}_2[\varphi(x_1,...,x_n,B_t-B_s)]|_{(x_1,...,x_n)=(B_{t_1},B_{t_2},...,B_{t_n})}]].
\end{eqnarray*}
The proof is complete now.\ \ \ \ \ \ \ \ \ \ \ \ \ \ \ \ \ \ \ \ \ \ \ \ \ \ \ \ \ \ \ \ \ \ \ \ \ \ \ \ \ \
\ \ \ \ \ \ \ \ \ \ \ \ \ \  $\blacksquare$\\  \\
\textbf{Acknowledgements} The authors thank Shige Peng, Ying Hu and
Mingshang Hu for careful reading and useful suggestions.\\\\
 \textbf{Reference}\\ \textbf{[1]}Artzner, Philippe; Delbaen, Freddy; Eber, Jean-Marc; Heath, David,
 Coherent measures of risk. Math. Finance 9 (1999), no. 3, 203--228. \\ \\
 \textbf{[2]}Barrieu, Pauline; El Karoui, Nicole, Optimal design of derivatives in illiquid markets. Quant. Finance 2 (2002), no. 3, 181--188.  \\\\
  \textbf{[3]}Barrieu, Pauline; El Karoui, Nicole, Optimal derivatives design under dynamic risk measures.
  Mathematics of finance, 13--25, Contemp. Math., 351, Amer. Math. Soc., Providence, RI, 2004.
\\\\
\textbf{[4]}Barrieu, Pauline; El Karoui, Nicole, Inf-convolution of
risk measures
 and optimal risk transfer. Finance Stoch. 9 (2005), no. 2, 269--298. \\\\
 \textbf{[5]}Chen, Zengjing; Epstein, Larry,
  Ambiguity, risk, and asset returns in continuous time. Econometrica 70 (2002), no. 4, 1403--1443.\\\\
   \textbf{[6]}El Karoui, N.; Peng, S.; Quenez, M. C.,
    Backward stochastic differential equations in finance. Math. Finance 7 (1997), no. 1, 1--71. \\\\
    \textbf{[7]}Rosazza Gianin, Emanuela,
Risk measures via g-expectations.
Insurance Math. Econom. 39 (2006), no. 1, 19--34.\\\\
\textbf{[8]}Delbaen, Freddy, Coherent risk measures,
 Cattedra Galileiana. [Galileo Chair] Scuola Normale Superiore, Classe di Scienze, Pisa, 2000.\\\\
  \textbf{[9]}Delbaen, Freddy, Coherent risk measures on general probability spaces. Advances in finance and stochastics, 1--37, Springer, Berlin, 2002.\\\\
\textbf{[10]}F\"{o}llmer, Hans; Schied, Alexander,
Convex measures of risk and trading constraints. Finance Stoch. 6 (2002), no. 4, 429--447.\\\\
 \textbf{[11]}F\"{o}llmer, Hans; Schied, Alexander, Robust preferences and
 convex measures of risk. Advances in finance and stochastics, 39--56, Springer, Berlin, 2002.\\\\
 \textbf{[12]}Frittelli, Marco; Rosazza Gianin, Emanuela, Putting order in
 risk measures. Journal of Banking and Finance
 26 (2002), no. 7, 1473-1486.\\\\
 \textbf{[13]}Frittelli, Marco; Rosazza Gianin, Emanuela, Dynamic Convex
 Risk measures, in: Risk measures for the 21st
 Century, G.Szeg\"{o}., J.Wiley, 2004, 227-248.\\\\
\textbf{[14]} Pardoux, Etienne; Peng, S. G.,
 Adapted solution of a backward stochastic differential equation. Systems Control Lett. 14 (1990), no. 1, 55--61.\\\\
\textbf{[15]} Peng, S., Backward SDE and related $g$-expectation.
Backward stochastic differential equations (Paris, 1995--1996), 141--159, Pitman Res. Notes Math. Ser., 364, Longman, Harlow, 1997.\\\\
\textbf{[16]}Peng,S., Dynamically consistent nonlinear evaluations
and expectations, arXiv:math/0501415v1 [math.PR] 24 Jan 2005.\\\\
\textbf{[17]}Peng,S., Multi-dimensional G-Brownian motion and
related stochastic calculus under G-expectation, Stochastic
Processes and their Applications, Vol. 118, Issue 12, Dec. (2008),
2223-2253.
\\ \\
\textbf{[18]}Peng, Shige, G-expectation, G-Brownian motion and
related stochastic calculus of It\^o type. Stochastic analysis and
applications, 541--567, Abel Symp., 2, Springer, Berlin, 2007.
\\ \\
\textbf{[19]}Peng,S., G-Brownian Motion and Dynamic Risk Measures
under Volatility Uncertainty, arXiv:0711.2834v1 [math.PR] 19 Nov
2007.

\end{document}